\documentclass[times, twocolumn]{aastex63}
\usepackage{graphicx, graphics}
\usepackage{CJK}
\usepackage{amsmath, amssymb, bm}
\usepackage[nodayofweek]{datetime}
\newcommand{\Ht}{H$_{2}$}
\newdateformat{monthyearday}{%
  \THEYEAR\ \monthname[\THEMONTH] \twodigit{\THEDAY}}

\published{\monthyearday\today}
\shortauthors{Blue Bird et al.}
\usepackage{romannum}
\usepackage[figuresright]{rotating}
\usepackage{afterpage} 
\usepackage{chngpage}
\usepackage{graphicx}
\usepackage[switch]{lineno}

\usepackage{hyperref}

\begin{document}
\pagenumbering{arabic}
\begin{CJK*}{UTF8}{gbsn}

\title{CHILES XI: Resolved \textsc{Hi} morphologies and \\
the evolution of the \Ht~/ \textsc{Hi} ratio over the last five billion years
}

\author{J. Blue Bird}\thanks{jbluebird.astro@gmail.com, \newline J. Blue Bird, formerly a Jansky Fellow of the NRAO. }
\affiliation{National Radio Astronomy Observatory, P.O. Box O, Socorro, NM 87801, USA}
\affiliation{Department of Astronomy, Columbia University, 550 West 120th Street, New York, NY 10027, USA}

\author{N. Luber}
\affiliation{Department of Astronomy, Columbia University, 550 West 120th Street, New York, NY 10027, USA}

\author{H. B. Gim}
\affiliation{Department of Physics, Montana State University, P.O. Box 173840, Bozeman, MT 59717, USA}

\author{J.H. van Gorkom}
\affiliation{Department of Astronomy, Columbia University, 550 West 120th Street, New York, NY 10027, USA}

\author{D.J. Pisano}
\affiliation{Department of Astronomy, University of Cape Town, Private Bag X3, Rondebosch 7701, South Africa}

\author{Min S. Yun}
\affiliation{Department of Astronomy, University of Massachusetts, Amherst, MA 01003, USA}

\author{E. Momjian}
\affiliation{National Radio Astronomy Observatory, P.O. Box O, Socorro, NM 87801, USA}

\author{K. M. Hess}
\affiliation{Department of Space, Earth and Environment, Chalmers University of Technology, Onsala Space Observatory, 43992 Onsala, Sweden}
and \affiliation{ASTRON, the Netherlands Institute for Radio Astronomy, Postbus 2, 7990 AA, Dwingeloo, The Netherlands}

\author{D. Lucero}
\affiliation{Department of Physics, Virginia Tech, 850 West Campus Drive, Blacksburg, VA 24061, USA}

\author{J. Donovan Meyer}
\affiliation{National Radio Astronomy Observatory (NRAO), 520 Edgemont Road, Charlottesville, VA 22903, USA}

\author{A. Chung}
\affiliation{Department of Astronomy, Yonsei University, 50 Yonsei-ro, Seodaemun-gu, Seoul 03722, Korea}

\begin{abstract}

We present the neutral gas morphology of four galaxies from $z$ = 0.22 to 0.47 obtained with the COSMOS \textsc{Hi} Large Extragalactic Survey (CHILES). The \textsc{Hi} is resolved at the highest redshift with the 7.5 $^{\prime\prime}$ beam of CHILES and 43 kpc linear scale, with all four galaxies having extended HI. Three are massive galaxies (M$_{*}$ $>$ $3\times10^{10}$ M$_{\odot}$), with \textsc{Hi} masses of 1.6 -- 6.7 $\times10^{10}$ M$_{\odot}$, and active star formation (3 -- 30 M$_{\odot}$ yr$^{-1}$). The morphology and kinematics of the galaxies vary from regular to disturbed, including an asymmetric \textsc{Hi} disk surrounding the fourth smaller galaxy (M$_{*}$ $\sim$ $10^{9}$ M$_{\odot}$). CO(1--0) observations of the sample, obtained with the LMT,  confirm the redshifts of three of the four galaxies and we derive \Ht~masses of 0.4 -- 5.2 $\times10^{10}$ M$_{\odot}$. JWST imaging with four combined NIRCam filters reveals disturbed stellar components with compact knots in two of the galaxies. We combine our new higher-redshift galaxies with previously published observations to conduct a more complete study of \textsc{Hi} and \Ht~evolution in the redshift range 0 -- 0.5. With our \textsc{Hi} flux-limited observations compared to similar lower redshift galaxies with high stellar mass (M$_{*}$ $>$ $10^{10}$ M$_{\odot}$), the results show the mean \Ht~/ \textsc{Hi} ratio at the highest redshift is 10.3 $\pm$ 3.4 larger than the mean \Ht~/ \textsc{Hi} ratio in the local Universe.

\vspace{5mm}
\vspace{5mm}

\end{abstract}
\section{Introduction}

Understanding the hydrogen gas content of galaxies is important for obtaining a full understanding of how galaxies grow and evolve. We know that cosmic star formation peaks at $z$ $\sim$ 2 and sharply declines to the present day \citep{Madau2014}, yet we know very little about the atomic neutral hydrogen (\textsc{Hi}) gas reservoirs surrounding individual galaxies that provide the initial fuel for star formation through redshifts beyond the local Universe ($z$ $>$ 0.06).

Star formation is well determined by numerous studies \citep{Hopkins2006, Madau2014}, whereas the gas components are more difficult to measure and are derived with different approaches, depending on the redshift. Our understanding of molecular gas in galaxies has improved with CO studies at redshift $z$ $>$ 1 \citep{Genzel2010, Tacconi2013, Scoville2017, Decarli2020, Walter2020}. In contrast, due to the faintness of the \textsc{Hi} 21 cm line, few direct \textsc{Hi} detections of individual galaxies have been made beyond the local universe with the exception of the BUDHIES survey of two galaxy clusters at 0.16 $<$ $z$ $<$ 0.22 \citep{Verheijen2007, Gogate2020}, the HIGHz Arecibo survey which measured massive galaxies in the redshift range 0.17 $<$ $z$ $<$ 0.25 \citep{Catinella2015}, and a single barred star-bursting spiral galaxy at $z$ = 0.38 \citep{Fernandez2016}. More recently, the MIGHTEE-HI survey \citep{Jarvis2025} find \textsc{Hi} emission in eleven galaxies at redshift 0.26 -- 0.38 using the MeerKAT telescope. In addition, the FUDS survey \cite{Xi2024} find five detections of \textsc{Hi} emission from galaxies at redshift 0.38 -- 0.40 with an angular resolution of $\sim$ 4$^\prime$ (1.3 Mpc) using the FAST telescope.			

With recently upgraded telescopes and SKA pathfinders, several studies have now published results on the average \textsc{Hi} content of galaxies out to $z$ = 1.4, using the stacking technique. The upgraded Giant Metrewave Radio Telescope (uGMRT) studies provide mass estimates from stacked \textsc{Hi} emission from blue, star-forming galaxies at higher redshift. \citet{Bera2019} stacked measurements indicate very little evolution to the present epoch in \textsc{Hi} gas content at $z$ $\sim$ 0.35, while \citet{Chowdhury2020} find an increase from this amount at $z$ $\sim$ 1 and \citet{Chowdhury2021} find a more significant increase at $z$ $\sim$ 1.3. Moreover, using CHILES (COSMOS \textsc{Hi} Large Extragalactic Survey) data, \citet{Luber2025a} find no evolution in the \textsc{Hi} gas content with stacked galaxies at intermediate stellar mass, through three redshift ranges $z$ = 0.15, 0.31, and 0.41. However, the average \textsc{Hi} mass fraction for high stellar mass galaxies appears to increase with redshift. These measurements are a good step in assessing the evolution of gas content with redshift by providing clues to average galaxy properties over large samples, but they cannot provide information about individual galaxies from direct imaging.

For the molecular hydrogen (\Ht) component, observational tracers (mainly CO lines and dust continuum) have been used to study \Ht~scaling relations of galaxies at higher redshifts \citep{Walter2020, Tacconi2020}. Combined detailed measurements of \textsc{Hi} and \Ht~exist in nearby galaxies \citep{Walter2020, Leroy2009, Catinella2018, Saintonge2017}, but to date there are only 18 galaxies with published detections in both neutral and molecular gas beyond the local Universe ($z$ $>$ 0.06). The follow-up CO observations \citep{Cortese2017} of massive $z$ $\sim$ 0.2 galaxies in the HIGHz survey \citep{Catinella2015} produce five galaxies with detections in both \textsc{Hi} and CO. With the BUDHIES survey \citep{Verheijen2007, Gogate2020} in two galaxy clusters around $z$ $\sim$ 0.2, the COOL BUDHIES \citep{Cybulski2016} follow-up produces six galaxies detected in both \textsc{Hi} and CO. From the CHILES survey, observations of a star-bursting spiral at $z$ = 0.38 \citep{Fernandez2016} show gas-rich measurements in both \textsc{Hi} and \Ht. In addition, a follow-up study of galaxy groups around $z$ $\sim$ 0.12 \citep{Hess2019} produces five galaxies (and one stacked) detected in both \textsc{Hi} and CO \citep{Hess2025}.

CHILES uses the NSF's Karl G. Jansky Very Large Array (VLA) to target the \textsc{Hi} line (1420 MHz rest frequency) over a continuous redshift range 0 $<$ $z$ $<$ 0.5 with 856 hours of observation in a single pointing in the COSMOS field. Over a frequency range of 1420 -- 964 MHz, CHILES produces resolved \textsc{Hi} images of galaxies with an angular resolution of 5 -- 7.6 $^{\prime\prime}$, respectively. The CHILES survey provides \textsc{Hi} content, morphology, and kinematics for a wide range of stellar masses and environments and a critical link between deep \textsc{Hi} stacking experiments over higher redshift ranges and \textsc{Hi} imaging experiments in specific, targeted galaxy environments.  

In this paper, we present four galaxies with \textsc{Hi} and \Ht~measurements drawn from \textsc{Hi} detections found in the CHILES catalog (Blue Bird et al. in preparation) with $z$ $>$ 0.2, M$_{*}$ $>$ $10^9$ M$_{\odot}$ and CO follow-up observations. These detections are gas-rich (M$_\mathrm{HI}$ = $1.6 - 6.7\times10^{10}$ M$_{\odot}$) star-forming galaxies (star formation rate (SFR) = 3 -- 30 M$_{\odot}$ yr$^{-1}$) with extended \textsc{Hi}. We obtain CO(1--0) observations from the Large Millimeter Telescope (LMT) of the four galaxies and derive \Ht~masses of 0.4 -- 5.2 $\times10^{10}$ M$_{\odot}$, for a combined look at the cold gas properties. With the advent of the James Webb Space Telescope (JWST), it is possible to shed light on stellar populations hidden behind dust in the distant Universe. We study morphology and kinematics using resolved images of the \textsc{Hi} gas and combine this with stellar morphology using images from the JWST. We compare this new higher-redshift sample with previous CHILES samples, providing for the first time a continuous look at the \textsc{Hi} + \Ht~gas properties out to a redshift of 0.5.

This paper is organized as follows. In Section 2, we discuss CHILES, LMT, SALT and JWST observations. In Section 3, we discuss the \textsc{Hi} search, sample, predictions, and galaxy properties. In Section 4, we present the \textsc{Hi} extent, morphology, and kinematics of four galaxies and also present JWST images. In Section 5, we present the \textsc{Hi} and \Ht~properties of a combined sample of CHILES galaxies that span a redshift range of 0.1 -- 0.5 and compare these with similar galaxies in the local Universe and around $z\sim$ 0.2. Section 6 provides a summary. Throughout this paper, we use J2000 coordinates, velocities in the optical convention, and a barycentric reference frame. This paper adopts a flat $\Lambda$CDM cosmology using H$_{o}$=70 km s$^{-1}$ Mpc$^{-1}$ and $\Omega_{M}$=0.3 to calculate distances and physical sizes.													

\section{Data}	

\begin{table}		
\begin{flushleft}		
\caption{CHILES Observations Over $z = 0.2 - 0.5$}		
\begin{tabular}{lrr}		
\hline		
\hline		
Field&	&	COSMOS\\
Right Ascension, Declination [deg]& 	&	150.35, 2.35\\
Array Configuration&	&	VLA -- B\\
Observation Date&	&	2013 -- 2019\\
Survey Epoch&	&	Epochs 1 -- 5\\
Integration [hours]&	&	856\\
Bandpass, Flux Calibrator&	&	3C286\\
Complex Gain Calibrator&	&	J0943 -- 0819\\[1.7ex]
\hline	
\hline
Redshift [$z$]&	0.2&	----\:\:\:\:\:\:0.5\\
Frequency [MHz]&	1184&	----\:\:\:\:\:964\\
Field of View [arcmin, Mpc]&	39, 7.7&	47, 17.2\\
Synthesized Beam [arcsec]$^{a}$&	6.1&	----\:\:\:\:\:\:7.6\\
Frequency Resolution [kHz]$^{b}$&	250&	----\:\:\:\:\:250\\
Velocity Resolution [km s$^{-1}$]$^{b}$&	63&	----\:\:\:\:\:\:\:78\\
Spatial Resolution [kpc]&	20&	----\:\:\:\:\:\:\:46\\
RMS Noise [$\mu \mathrm{Jy}$ bm$^{-1}$ ch$^{-1}$] $^{c}$&	26&	\:\:\:\:\:\:\:45\\
1$\sigma$ N$_{HI}$ [cm$^{-2}$] $^{c}$&	6.8$\times10^{19}$&	1.6$\times10^{20}$\\[1.7ex]
\hline		
\hline		
\multicolumn{3}{p{8cm}}{\footnotesize{$^{a}$Average of Bmaj$\times$Bmin. $^{b}$After Hanning smoothing plus additional velocity smoothing. $^{c}$Parameters refer to $z$ = 0.2 and $z$ = 0.5 specifically.} }
\end{tabular}		
\end{flushleft}		
\label{tab_obs}		
\end{table}	

\subsection{CHILES Observations} \label{sec_chiles}	    
CHILES data reduction is carried out in CASA \citep{CASA2022} and the data are calibrated in a standard manner with some pipeline modifications using the statistical flagging task in CASA, \texttt{rflag}, and using masks to blank out the bad RFI frequency ranges, interpolating across them when applying the calibration (Pisano et al. in preparation). The data are imaged with two separate methods in order to explore the most effective way to remove side-lobes caused by outlying continuum sources that are beyond the first null of the VLA primary beam. A key difference in the methods is how these out-of-field sources are modeled and subtracted. See \citet{Dodson2022} and \citet{Luber2025a} for detailed explanations of the two methods. The baselines at the locations of individual \textsc{Hi} detections are fit with a first-order polynomial. In this work, we focus on a subset of the overall CHILES frequency range that covers 1184 to 964 MHz, which corresponds to $z$ = 0.2 -- 0.5. See Table \ref{tab_obs} for observation details. The ranges shown for typical root mean square (RMS) noise and column density are within theoretical estimates.	

\begin{figure*}
\includegraphics[scale=0.85]{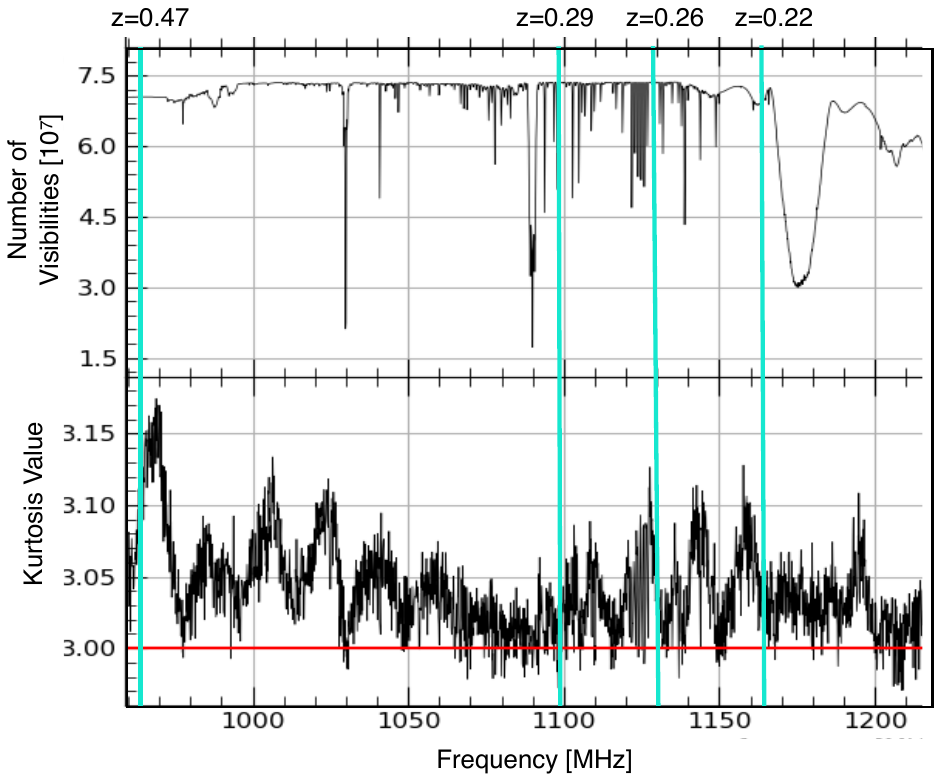}
\centering
\caption{
Characterization of the locations of \textsc{Hi} detections within the CHILES image cubes. \textbf{Top:} The number of unflagged visibilities used in imaging the CHILES data. \textbf{Lower:} Kurtosis values of the CHILES data, following the processing outlined in \citet{Luber2025a}. A kurtosis value of three indicates perfectly Gaussian data. The kurtosis values indicate that the noise deviates from Gaussianity in frequency ranges 964-1060 MHz and 1100-1200 MHz. The four \textsc{Hi} detections (turquoise lines) are in frequency ranges where little flagging is done (top) and where the noise is closer to Gaussian (bottom).
}
\label{fig_rfi}
\end{figure*}

The CHILES survey is designed to detect a galaxy with a \textsc{Hi} mass of $3\times10^{10}$ M$_{\odot}$ at its highest redshift of $z$ = 0.5. In addition to the redshift range 0.10 $< z <$ 0.2 heavily impacted by RFI \citep{Hess2019, Hess2025}, we also find that detecting 21-cm \textsc{Hi} emission is proving to be challenging at higher redshifts beyond $z \sim$ 0.2. Since the line is faint, long integration times are needed, bringing up side-lobes from continuum sources far away from the field center, beyond the first null of the primary beam, where calibration is imperfect. RFI also complicates calibration and can cause non-flat spectral baselines. There is a risk that the methods used to remove artifacts and subtract continuum also remove faint \textsc{Hi} signals. In Figure \ref{fig_rfi}, we show the amount of data flagged at different frequencies (top) and a measure of the non-Gaussianity of the noise (bottom). The horizontal red line with a kurtosis value of three indicates the value for perfectly Gaussian data, with the expected variation of the CHILES data being 0.004. Known bright side-lobe patterns over this redshift range, caused by outlying continuum sources that are beyond the first null of the VLA primary beam, inject the image plane with pixel values that significantly deviate from an expected Gaussian noise distribution and present as structured image-plane features \citep{Luber2025a}. The four detections discussed in this paper are all at frequencies where little flagging is done and the noise is closer to Gaussian.

\subsubsection{Identification of \textsc{Hi} Sources} \label{sec_search}
We describe the search for \textsc{Hi} in the CHILES data in the redshift range $z$ = 0.2 -- 0.5 in a set of 12 image cubes, each of which spans 24 MHz with 250 kHz channels. We perform two types of search, untargeted and targeted. Each type of search is carried out first with visual inspections and then with the Source Finding Application (SoFiA) \citep{Serra2015,Westmeier2021} suite of semi-automated algorithms.

We begin the untargeted search with visual inspection in the center of the field of view and work around the center in $15^\prime\times15^\prime$ subsections moving outward just beyond the Half Power Beam Width (HPBW), which spans a linear scale of 7.7 -- 17.2 Mpc at $z$ = 0.2 -- 0.5. Promising candidates are then cross-checked for optical counterparts. We visually detect \textsc{Hi} in galaxy 1203414 at $z$ = 0.26 (featured in this analysis), and an additional galaxy which will be featured in a future low-mass galaxy analysis. Untargeted searches using the SoFiA reliability function \citep{Serra2015} in this redshift regime still require quite a bit of manual interaction and do not lead to any detections.

We start the targeted search at the locations of galaxies with the highest predicted \textsc{Hi} masses. We estimate the \textsc{Hi} mass using the relation of the linear combination of stellar surface density and NUV-r color of xGASS galaxies \citep{Catinella2010,Catinella2012,Catinella2018} (and references therein). Using k-corrected G10/COSMOS catalog values \citep{Davies2015}, we obtain \textsc{Hi} mass estimates for 1699 star-forming galaxies (NUV-r $<$ 3) with spectroscopic redshifts, over the CHILES redshift range $z$ = 0 -- 0.5. The number of galaxies that have all the parameters used in the relation limits the number of galaxies with mass estimates. Moreover, 1226 of these galaxies have redshifts $z$ $>$ 0.2, and 331 galaxies have \textsc{Hi} mass estimates $>$ $1\times10^{10}$ M$_{\odot}$. There are 45 galaxies with a predicted mass greater than $3\times10^{10}$ M$_{\odot}$ and 11 galaxies with a predicted mass greater than $5\times10^{10}$ M$_{\odot}$. The visual and SoFiA targeted searches at these locations result in zero detections. The \textsc{Hi} mass detection limit (3$\sigma$) for CHILES (assuming a line width of 150 km s$^{-1}$) at the edge of the HPBW at $z$ = 0.2 is $1\times10^{10}$ M$_{\odot}$.

We conduct targeted searches, using SoFiA, of the remaining galaxies with predicted \textsc{Hi} mass $<$ $10^{10}$ M$_{\odot}$, and of the remaining star-forming galaxies in the G10/COSMOS catalog \citep{Davies2015}. Detailed descriptions of the use of SoFiA can be found in \citet{BlueBird2020}. For all of our searches, a noise scaling filter is applied along the velocity axis to normalize the cube by the noise level per channel to account for variable noise characteristics throughout the cube. The Smooth and Clip algorithm \citep{Serra2012} smooths the image cubes at several angular resolutions, and we use Gaussian kernels that vary from no smoothing to 1.5, 2, and 3 times the synthesized beam that varies with redshift. Similarly, image cubes are smoothed to different velocity resolutions, varying from no smoothing to 3, 7 and 9 times the channel width using boxcar kernels. A specified relative flux threshold of 2.5 -- 3.5 (in multiples of the noise level) is applied to the spatial and velocity resolutions, to extract and mark the significant pixels on each scale. We run some searches using mask dilation, which iteratively grows the original mask outward in all three dimensions until the integrated flux within the mask does not increase any further. This can help to pick up the faint parts of sources that are below the detection threshold. A combination of these settings contributes to finding three detections that are featured in this analysis; Galaxies 1432541 and 1421092 (star-forming galaxies with no predicted \textsc{Hi} masses) and 1199583 (which has a predicted \textsc{Hi} mass $<$ $10^{10}$ M$_{\odot}$).

\begin{table*}								
\caption{Galaxy Properties}								
\begin{adjustwidth}{-0.1cm}{0cm}								
\centering								
\begin{tabular}{lllllllll}								
\hline								
\hline
COSMOS&	R.A.&	Dec.&	\textsc{Hi}, \Ht~(optical)&	\textsc{Hi} Freq.&	SFR (SFR$_\mathrm{1.4}$)&	NUV-r&	$\mu_{*}$ $10^{8}$&	Incl.\\
2008 ID&	[J2000]&	[J2000]&	redshift&	[MHz]&	[M$_{\odot}$/yr]& 	&	[M$_{\odot}$/kpc$^{2}$]&	[deg]\\
(1)&	(2)&	(3)&	(4)&	(5)&	(6)&	(7)&	(8)&	(9)\\
\hline								
1432541&	150.2591&	2.5281&	0.4703, 0.4729* (0.4712)&	966.04&	29.9 (36.5)&	2.7&	NA&	67\\
1199583&	150.2709&	2.4429&	0.2922, NA (0.2918)&	1099.23&	3.9 (NA)&	0.4&	0.12&	75\\
1203414&	150.3374&	2.4125&	 0.2577, 0.2577 (0.2577)&	1129.33&	2.5 (6.6)&	2.7&	1.15&	45\\
1421092&	150.2826&	2.6103&	0.2194, 0.2202 (0.2198)&	1164.85&	5.8 (13.1)&	1.8&	NA&	79\\[1.7ex]
\hline								
\hline	
COSMOS&	D$_\mathrm{HI}$ (D$_\mathrm{opt}$)&	M$_{*}$ $10^{10}$&	M$_\mathrm{HI}$ $10^{10}$&	M$_\mathrm{H2}$ $10^{10}$&	W$_\mathrm{HIint}$&	W$_\mathrm{H2int}$&	PA$_\mathrm{HI}$&	S/N$_\mathrm{HI}$\\	2008 ID&	[kpc]&	[M$_{\odot}$]&	[M$_{\odot}$]&	[M$_{\odot}$]&	[km/s]&	[km/s]&	[deg]&	[$\mathrm{N}$ ($\mathrm{Pk}$)]\\
((10)&	(11)&	(12)&	(13)&	(14)&	(15)&	(16)&	(17)&	(18)\\
\hline								
1432541&	96$\pm$43 (57)&	19.30$\pm$2.70&	6.68$\pm$1.74&	5.18$\pm$1.64&	772$\pm78$&	718$\pm120$&	61&	6.1 (5.1)\\
1199583&	72$\pm$29 (19)&	0.17$\pm$0.02&	2.34$\pm$0.75&	$<$0.27&	334$\pm68$&	NA&	88&	3.5 (4.1)\\
1203414&	118$\pm$26 (51)&	7.14$\pm$0.68&	2.80$\pm$0.42&	0.44$\pm$0.27&	527$\pm66$&	205$\pm102$&	256&	4.3 (6.0)\\
1421092&	55$\pm$22 (41)&	2.69$\pm$0.86&	1.59$\pm$0.65&	0.63$\pm$0.55&	316$\pm64$&	498$\pm99$&	33&	3.0 (3.2)\\[1.7ex]
\hline								
\hline								
\multicolumn{9}{p{17.5cm}}{\footnotesize{(1) COSMOS 2008 ID;  (2) Units of right ascension are in degrees;  (3) Units of declination are in degrees;  (4) \textsc{Hi} redshifts are from CHILES, \Ht~redshifts are from LMT observations, and optical redshifts are from the G10/COSMOS v05 catalog;  (5) \textsc{Hi} frequency;  (6) SFR, from SED fitting on UV through IR;  (7) NUV-r, using GALEX NUV magnitude and SDSS DR7 r-band magnitude;  (8) Stellar surface density;  (9) Inclination is calculated such that 0$^{\circ}$ is face-on; (10) COSMOS 2008 ID;  (11) \textsc{Hi} diameter, along the \textsc{Hi} major axis, corrected for the beam smearing; Optical diameter, along the SDSS r-band isophotal major axis;  (12) Stellar mass, from SED fitting on UV through IR;  (13) \textsc{Hi} mass, corrected for the primary beam;  (14) \Ht~mass, using a CO conversion factor of 4.3;  (15) \textsc{Hi} integrated line width, corrected for line broadening;  (16) \Ht~integrated line width, corrected for line broadening;  (17) \textsc{Hi} kinematic position angle, along the \textsc{Hi} major axis;  (18) Signal-to-noise ratio, using the lowest column density contour of the \textsc{Hi} intensity map (Equation 4); Signal-to-noise ratio, using \textsc{Hi} peak flux (Equation 3). *Redshift is for both \Ht~from LMT observation and \textsc{[Oii]} from SALT observation.} }
\end{tabular}								
\label{tab_prop}								
\end{adjustwidth}								
\end{table*}			

\subsubsection{\textsc{Hi} Sample} \label{sec_sample}
Our sample of galaxies in this paper is drawn from \textsc{Hi} detections found in the CHILES catalog (Blue Bird et al. in preparation). The \textsc{Hi} detections presented in this study consist of four galaxies in the redshift range of 0.22 $<$ $z$ $<$ 0.47 with M$_{*}$ $>$ $10^9$ M$_{\odot}$. It should be noted that because we are using \textsc{Hi} and CO in emission, we are necessarily biased towards star-forming, gas-rich objects.  

Instead of relying on the spectrum to confirm the detection of \textsc{Hi}, we rely on the presence of coherent contours in the channel maps. \textsc{Hi} confirmation incorporates the criteria outlined in the CHILES pilot study \citep{Fernandez2013}. This includes having spatially coherent emission in the immediate vicinity of the galaxy, which is seen in at least three consecutive channels, with at least 3$\sigma$ in two of them, and within 1 MHz (314 -- 458 km s$^{-1}$ over $z$ = 0.2 -- 0.5)  of the frequency corresponding to the optical redshift. To support the significance, these detections coincide in sky and frequency using G10/COSMOS spectroscopic redshifts, Hubble Space Telescope (HST) I-band mosaic images \citep{Koekemoer2007}, and JWST imaging \citep{Casey2023}. Moreover, the channels with coherent contours match in two sets of cubes imaged using two different continuum subtraction methods \citep{Dodson2022,Luber2025a}.

\subsubsection{\textsc{Hi} Properties} \label{sec_prop}
The \textsc{Hi} properties of the four galaxies are listed in Table \ref{tab_prop}. The galaxies are referred to using their COSMOS 2008 ID. 

\textsc{Hi} mass is calculated following Equation 48 from \citet{Meyer2017}.

\begin{equation} 
\begin{split}
\left( \frac {M_{\mathrm{HI}}}{\mathrm{{M_{\odot}}}} \right) = 
49.7 \times
\left( \frac{D_{\mathrm{L}}}{\mathrm{Mpc}} \right)^2 
\left( \frac{S}{\mathrm{Jy~Hz}} \right) 
\end{split}
\end{equation}

where $D_{\text{L}}$ is the luminosity distance, and S is the flux integral over the source mask. The \textsc{Hi} masses have been corrected for the primary beam. The RMS noise is measured off to the side of the galaxy over the same sized area enclosed in the source mask.

\textsc{Hi} column density is calculated following Equation 76 from \citet{Meyer2017}.

\begin{equation} 
\begin{split}
\left( \frac {N_{\mathrm{HI}}} {\mathrm{cm}^{-2} } \right) = 
2.33 \times 10^{20} \:\: (1+z)^{4} \\ \times
\left( \frac{ab}{\mathrm{arcsec^{2}}} \right)^{-1}
\left( \frac{S}{\mathrm{Jy~Hz}} \right)  
\end{split}
\end{equation}

where $z$ is the redshift, $a$ and $b$ are the FWHM of the major and minor axes of the synthesized beam in arcsec, and S is the flux.

Signal-to-noise ratio (SNR) for \textsc{Hi} is calculated in two ways and is listed in Table \ref{tab_prop}. First, the peak SNR is determined using Equation 153 from \citet{Meyer2017}.

\begin{equation} 
S/N_{\text{ peak}} = \frac{S_{\text{peak}}} {\sigma_{\text{chan}}} 
\end{equation} 

Second, for extended sources, the SNR is determined from the column density sensitivity using the lowest contour shown in the \textsc{Hi} intensity maps, following Equation 158 from \citet{Meyer2017}. The solid angle covered by the beam $\Omega_{\mathrm{bm}}$ = $\pi$ $a$$b$ / 4 ln 2 and the channel RMS is $\sigma_{\text{chan}}$.

\begin{equation} 
\begin{split}
S/N_{\mathrm{\:N_{HI}}}  = 
5.50 \times 10^{-23} \:\: 
\left( {1+z} \right)^{\frac{2}{7}} 
\left( \frac{N_{\mathrm{HI}}} {\mathrm{cm^{-2}}} \right) \\ \times
\left( \frac{\Omega_{\mathrm{bm}}} {\mathrm{arcsec^{2}}} \right)
\left( \frac{\Delta V_{\rm rest}} {\mathrm{km~s^{-1}}} \right)^{-\frac{1}{2}}  
\left( \frac{{\Delta \nu_{\rm chan}}} {\mathrm{Hz}} \right)^{-\frac{1}{2}}  
\left( \frac{\sigma_{\text{chan}}} {\mathrm{Jy}} \right)^{-1} 
\end{split}
\label{eq_nh1}
\end{equation}

Additional \textsc{Hi} properties are listed in Table \ref{tab_prop}. The PA is taken to match the \textsc{Hi} kinematic major axis of the velocity field. The \textsc{Hi} diameter D$_{HI}$ is measured along the \textsc{Hi} major axis at a limiting column density of $1.25\times10^{20}$ cm$^{-2}$ (1 M$_{\odot}$ pc$^{-2}$) \citep{Rhee1997}. D$_{HI}$ is corrected for beam smearing effects using a Gaussian approximation \citep{Wang2016}:

\begin{equation} \label{eqd}
D_{\mathrm{HI}}  = \sqrt{ (D_{\mathrm{HIo}}^2) - (B^2) }
\end{equation}
where $D_{\text{HI}}$ and $D_{\text{HIo}}$ are the corrected and uncorrected \textsc{Hi} diameters, and B is the synthesized beam.

W$_{\text{int}}$ is the emission line width over which the global \textsc{Hi} profile is integrated, taken from the maximum velocity of the rising and declining parts of the \textsc{Hi} position-velocity (PV) diagram in frequency. W$_{\text{int}}$ is corrected for line broadening using:

\begin{equation} \label{eqw}
W_{\text{HI}}  = \sqrt{ (W_{\text{obs}}^2) - (W_{\text{res}}^2) }
\end{equation}

The emission line widths in the observed frame are given in frequency, $\nu_{\text{obs}}$, while those in the source rest frame are specified in terms of velocity, $\Delta V_{\text{rest}}$. This is calculated using Equation 20 from \citet{Meyer2017}.

\begin{equation} 
\left( \frac{\Delta V_{\text{rest}}} {\mathrm{km~s^{-1}}} \right) = 
\frac{\Delta {\nu_{\text{obs}}}} {{\nu_{\text{obs}}}} \times c 
\end{equation}

Stellar properties of the galaxies come from the extensive multi-wavelength data available from ancillary observations in the COSMOS field \citep{Scoville2007}. Stellar mass and SFR are estimated from SED fitting using a 38-band panchromatic photometry approach spanning from the far-ultraviolet to the far-infrared from the G10/COSMOS catalog with the MAGPHYS energy-balanced SED fit program \citep{DaCunha2008} and are shown in Table \ref{tab_prop}. The uncertainty in the stellar mass of the four galaxies presented in this study is on the order of 15\% and the SFR uncertainty is on the order of 12\%. The 1.4 GHz \citep{Gim2025} SFR is calculated using Equation 10 of \citet{Murphy2012} and is listed in Table \ref{tab_prop}.

\begin{figure}[h]
\includegraphics[scale=0.27]{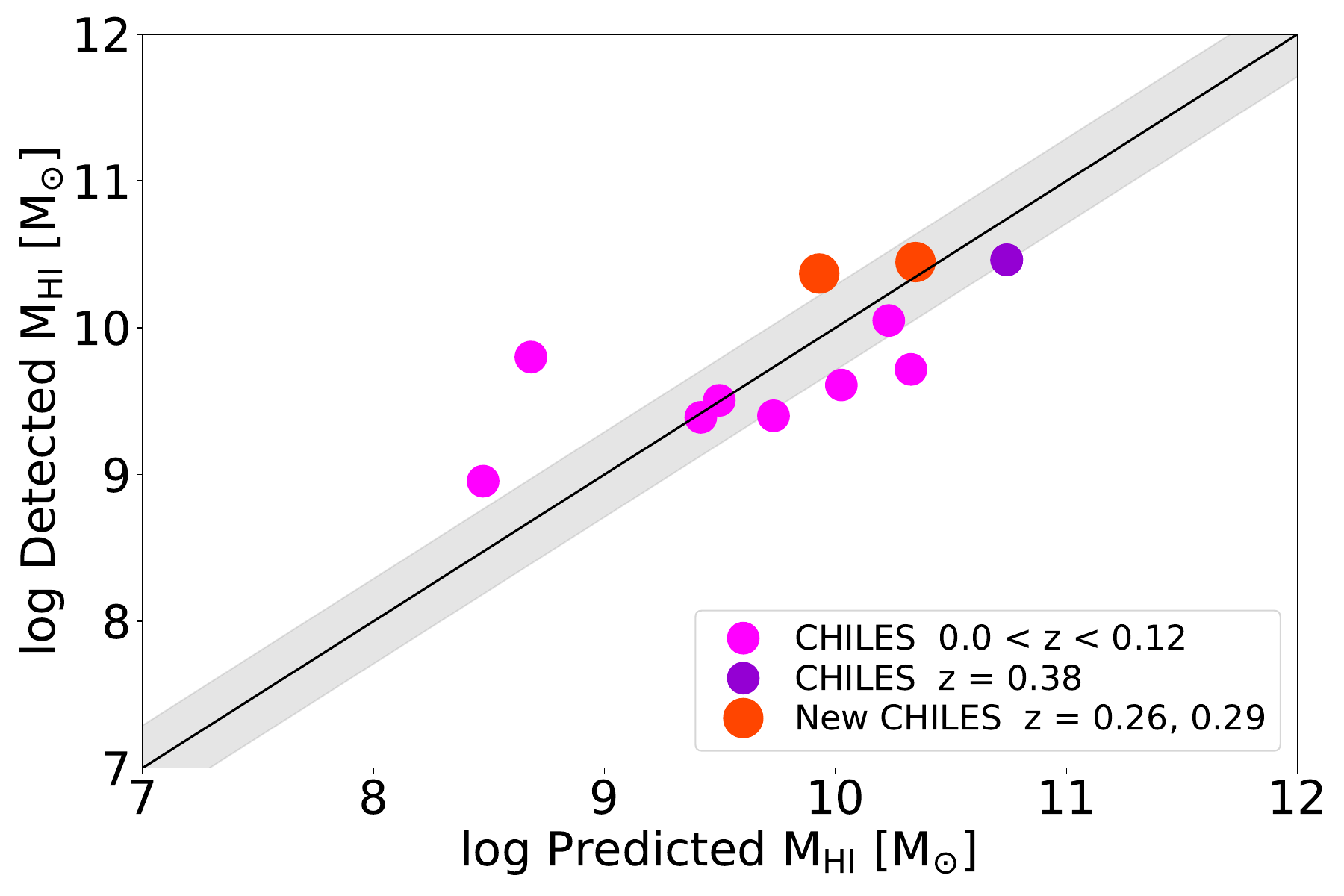}
\caption{
Relation between the predicted \textsc{Hi} mass and the detected \textsc{Hi} mass. The pink circles are CHILES galaxies from $z$ = 0 -- 0.12 \citep{BlueBird2020, Hess2019, Hess2025}, the purple circle is the CHILES galaxy at $z$ = 0.38 \citep{Fernandez2016}, and the orange circles are new galaxies presented in this work. The diagonal line indicates equal masses, with the dex scatter of the \citet{Catinella2012} predicted mass relation shown in the shaded region.
}
\label{fig_pred}
\end{figure}

\subsubsection{\textsc{Hi} Predictions}
In Figure \ref{fig_pred}, we compare the detected \textsc{Hi} mass and the predicted \textsc{Hi} mass for CHILES galaxies that have the parameter space to calculate the predicted mass using the relation from \citet{Catinella2012}. This includes the galaxy at $z$ = 0.38 from \citet{Fernandez2016}, four out of 10 galaxies at $z$ $<$ 0.1 from \citet{BlueBird2020}, four out of the 16 galaxies at $z$ $\sim$ 0.12 from \citet{Hess2019, Hess2025}, and two out of the four new galaxies presented in this work. The stellar mass and NUV-r information comes from the G10/COSMOS catalog. At high redshift, there are a limited number of galaxies that have z-band radius information (used to calculate the stellar mass surface density). Therefore, the HST I-band (F814W) radius \citep{Tasca2009} is used instead. The I-band should only differ from the z-band half-light effective radius by 3\% on average for galaxies between $z$ = 0.01 -- 0.1 \citep{Lange2015}. Moreover, as mentioned in Section \ref{sec_search}, the number of galaxies that have all the parameters used in the relation limits the number of galaxies with mass estimates. The relation in \citet{Catinella2012} is calibrated using a sample of galaxies with stellar mass greater than $10^{10}$ M$_{\odot}$, and might be less reliable for smaller galaxies. We find that our \textsc{Hi} detections generally agree with the predicted \textsc{Hi} masses. The two outliers farthest from the line are two dwarf galaxies (M$_{*}$ $<$ $10^{9}$ M$_{\odot}$). 

Although our detected \textsc{Hi} masses agree quite well with the predicted \textsc{Hi} masses, we have a large number of non-detections with upper limits well below the predictions. For example, assuming the local \textsc{Hi} mass function \citep{Martin2010}, we should have found 12 galaxies with larger \textsc{Hi} mass than our highest redshift \textsc{Hi} detection in the redshift range $z$ $>$ 0.3 and using the \citet{Catinella2012} prediction, six galaxies should have had a higher \textsc{Hi} mass in that range. The reasons for the non-detections are explored in a following paper and are beyond the scope of this paper.

\begin{figure*}
\includegraphics[scale=0.65]{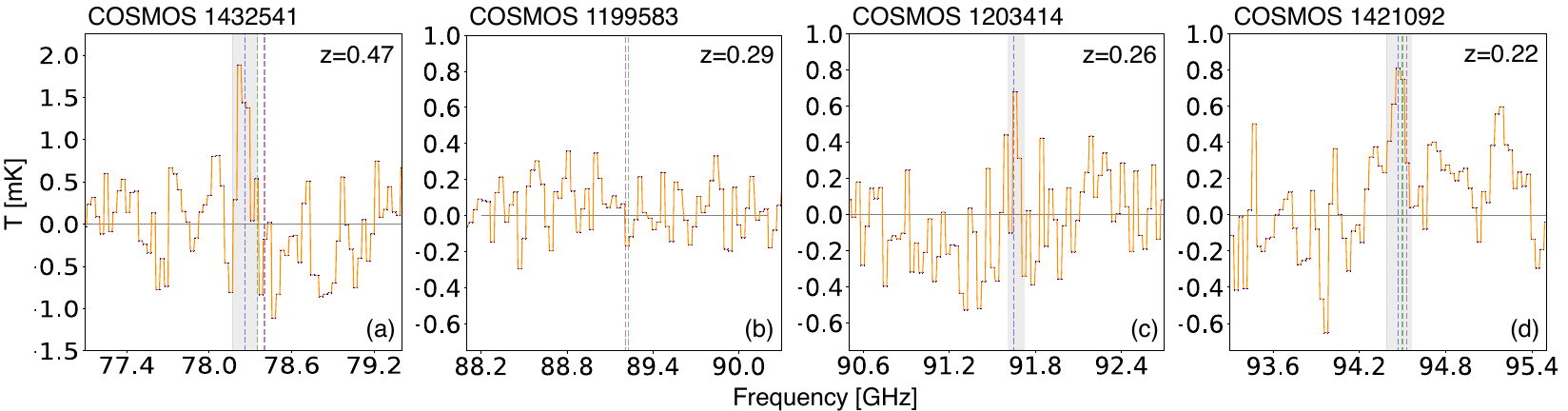}
\caption{
CO (1--0) spectra obtained using the Redshift Search Receiver (RSR) on the LMT. The shaded gray area indicates the area of detection. The CO line center frequency is shown (blue dashed line) along with the CO-rest-frequency equivalents for the \textsc{Hi} line center (purple dashed line) and the optical redshift (green dashed line). \textbf{Panel a:} CO(1--0) detection for Galaxy 1432541 at 78.3 GHz ($z$ = 0.4729). \textbf{Panel b:} CO(1--0) non-detection for galaxy 1199583. \textbf{Panel c:} CO(1--0) detection for galaxy 1203414 at 91.7 GHz ($z$ = 0.2577). \textbf{Panel d:} CO(1--0) detection for galaxy 1421092 at 94.5 GHz ($z$ = 0.2202).
}
\label{fig_lmt}
\end{figure*}	

\begin{table*}								
\caption{LMT Observations}								
\begin{adjustwidth}{-1.5cm}{0cm}								
\centering								
\begin{tabular}{lllllllll}								
\hline								
\hline								
COSMOS&	CO&   	CO Freq.&	Obs.&	Spatial Res.&	Spectral Res.&	RMS Noise& 	SN&	Lco (det)\\
2008 ID&	redshift&	[GHz]&	[hours]&	[arcsec, kpc]&	[km/s]&	[mK]&	$\mathrm{Pk}$&	$10^{9}$ [K km s$^{-1}$ pc$^{2}$]\\
(1)&	(2)&	(3)&	(4)&	(5)&	(6)&	(7)&	(8)&	(9)\\
\hline								
1431541&    	0.4729&	78.35&	1.58&	18, 106&	120&	0.40&  	4.5&	12.05$\pm$3.82 \\
1199583&	NA&	89.23&	4.65&	16, 69&	105&	0.15&   	NA&	$<$0.63$^{a}$\\
1203414&	0.2577&	91.65&	1.50&	15, 61&	102&	0.26& 	2.4&	1.01$\pm$0.63\\
1421092&	0.2202&	94.50&	1.58&	15, 53&	99&	0.26& 	2.8&	1.46$\pm$1.28\\ [1.7ex]
\hline								
\hline								
\multicolumn{9}{p{15.1cm}}{\footnotesize{(1) COSMOS 2008 ID;  (2) CO(1-0) line redshift;  (3) CO(1-0) line frequency;  (4) Total observation hours;  (5) Spatial resolution for 50 m dish;  (6) Spectral resolution at 32 MHz;  (7) RMS noise; (8) Signal-to-noise ratio, using CO peak flux (Equation 3).  (9) CO(1--0) line luminosity. $^{a}$Value is a 3$\sigma$} upper limit.}								
\end{tabular}								
\label{tab_lmt}								
\end{adjustwidth}								
\end{table*}								

\subsection{LMT Observations} \label{sec_lmt}
The LMT observations on 2023 January 25 to February 14 use a total time of 9.31 hours for all four targets [LMT 2023-S1-US-18, PI: J. Blue Bird]. The expected luminosities of the CO line are determined from UV + IR SFR measurements from the G10/COSMOS catalog (\citep{Davies2015} and references therein) for the target sample following the well-known correlation from \citet{Solomon2005}. The detected luminosities L$_{co}$ are listed in Table \ref{tab_lmt}. The LMT RSR calibration adopts the standard vane cal using a room-temperature black body load, and utilizes \textit{dreampy} LMT reduction software. The unfitted spectrum is the raw ON-OFF spectrum averaged for the two spatial beams and two polarizations. This ON-OFF differencing is done electronically at 1 kHz frequency, to keep the baseline flat. When averaged together with the 1/$\sigma$$^2$ weight, the final spectrum of the target has an RMS noise range of $\sigma$ = 0.15 -- 0.40 mK. The spectral resolution has a fixed value of 31 MHz over the entire frequency band, resulting in a rest-frame resolution of 99 -- 120 km s$^{-1}$. The 50 m LMT has a frequency-dependent beam size (HPBW) ranging from 15 -- 18 arcsec for these detections. This translates to a linear scale of 53 -- 106 kpc, which is lager than the optical diameter for each galaxy, likely avoiding attenuation of the CO emission.

\begin{figure}
\centering
\includegraphics[scale=0.8]{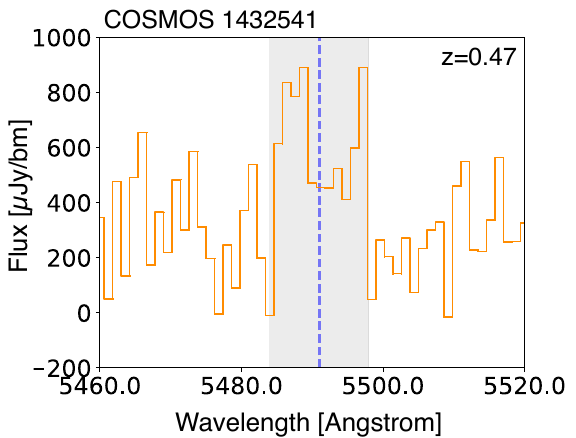}
\caption{
\textsc{[Oii]} a long-slit spectrum for galaxy 1432441 obtained using SALT. The shaded gray area indicates the area of detection. The \textsc{[Oii]} doublet center wavelength is shown (blue dashed line), which coincides with the observed CO(1-0) line center for this galaxy interpreted as $z = 0.473$.
}
\label{fig_salt}
\end{figure}

\begin{figure*}
\centering
\includegraphics[scale=0.65]{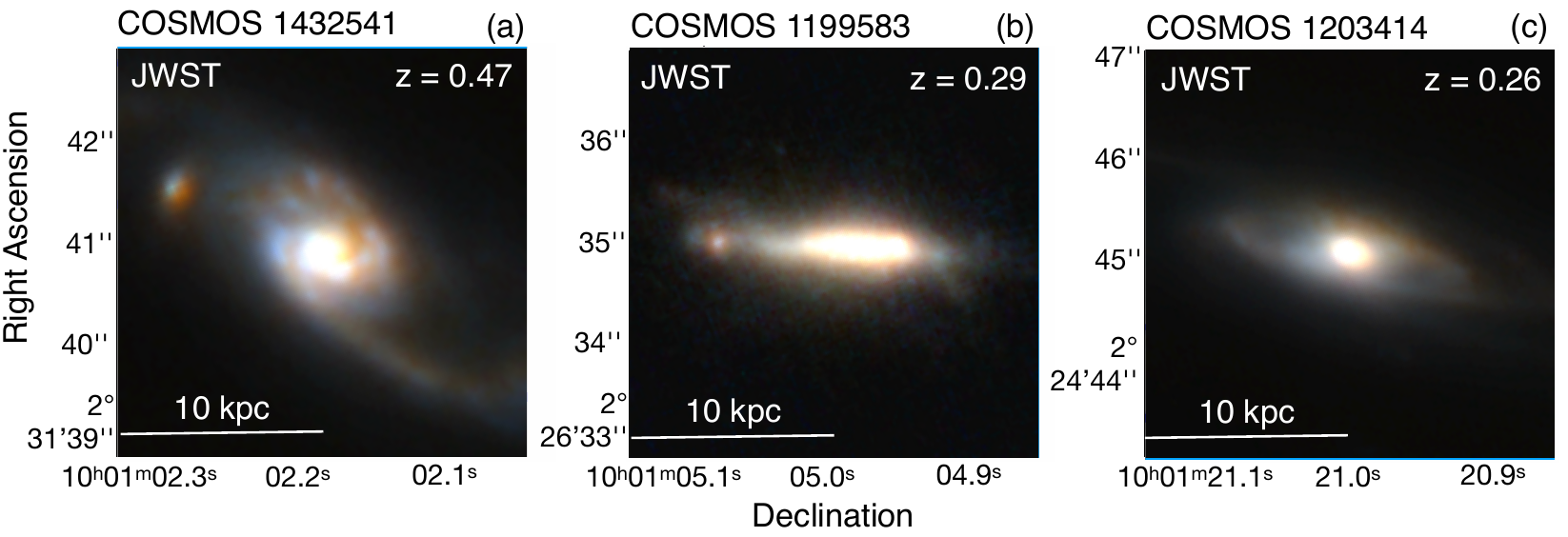}
\caption{
Near-IR JWST images. These images are made by combining four filters (1.15, 1.50, 2.77, 4.40 $\mu m$) from the COSMOS-Web NIRCam survey \citep{Casey2023}. The filters are assigned to the red, green, and blue channels as (R,G,B) = (F444W + F277W, F277W + F150W,  F150W + F115W). RGB images are constructed according to the Trilogy9 prescription \citep{Coe2012}. There is no JWST image for Galaxy 1421092 as it falls out of the COSMOS-Web FOV. \textbf{Panel a:} JWST image for galaxy 1432541 at $z$ = 0.47.  \textbf{Panel b:} JWST image for galaxy 1199583 at $z$ = 0.29. \textbf{Panel c:} JWST image for galaxy 1203414 at $z$ = 0.26. 
}
\label{fig_jwst}
\end{figure*}

The LMT observations result in three CO(1--0) detections and one upper limit, shown in Figure \ref{fig_lmt} and summarized in Table \ref{tab_lmt}. For galaxy 1432541, a bright line is detected at 78.3 GHz (peak SNR = 4.5) and is interpreted as the CO(1--0) transition at $z$ = 0.4729. This CO redshift agrees well with our \textsc{[Oii]} redshift (Figure \ref{fig_salt}), but it is slightly larger than the \textsc{Hi} redshift of 0.4703. The CO integrated line width is 718 $\pm$ 120 km s$^{-1}$, with the center offset from the \textsc{Hi} line center ($\pm$ 78 km s$^{-1}$) by -526 km s$^{-1}$ and from the optical redshift by -352 km s$^{-1}$. The spatially resolved \textsc{Hi} emission shows a nearly 100 kpc diameter disk-like structure while the JWST/NIRCam image shows a clearly disturbed stellar morphology, and the observed differences in velocity may reflect spatially distinct kinematics of these line tracers (see further discussions in \ref{im_1}). Galaxy 1199583 is not detected in CO(1--0), and a 3$\sigma$ upper limit is calculated for the \Ht~measurement. A single line is detected for galaxy 1203414 at 91.7 GHz (peak SNR = 2.4), interpreted as $z$ = 0.2577. The CO integrated line width is 205 $\pm$ 102 km s$^{-1}$, with the center aligned by -8 km s$^{-1}$ with the \textsc{Hi} line center ($\pm$ 66 km s$^{-1}$) of our highest \textsc{Hi} SNR detection and its optical redshift. A single line is detected for galaxy 1421092 at 94.5 GHz (peak SNR = 2.8), interpreted as $z$ = 0.2202. The shaded gray area indicates the area of detection and does not include the adjacent lower-frequency broader base. The CO integrated line width is 488 $\pm$ 99 km s$^{-1}$, with the center offset from the \textsc{Hi} line center ($\pm$ 64 km s$^{-1}$) by -194 km s$^{-1}$ and from the optical redshift by -96 km s$^{-1}$. 

\Ht~mass is listed in Table \ref{tab_prop} and is calculated using Equation 3 from \citet{Solomon2005}, assuming $\alpha_{\text{CO}} =$ 4.3. The conversion factor includes a correction for heavy elements, such as helium, of 36\% typically assumed for the Milky Way \citep{Bolatto2013, Saintonge2022}.

\begin{equation} 
\begin{split}
\left( \frac {M_{\mathrm{H2}}}{\mathrm{{M_{\odot}}}} \right) = 
3.25 \times 10^{7} \:
(1+z)^{-3} \: \\ \times
\left( \frac{D_{\mathrm{L}}}{\mathrm{Mpc}} \right)^2 
\left( \frac{S_{\text{CO}}{\Delta v_{}} }{\mathrm{Jy~km~s^{-1}}} \right) \:\:
\left( \frac{{\nu_{\rm obs}}} {\mathrm{GHz}} \right)^{-2} \:\:
\alpha_{\text{CO}}
\end{split}
\end{equation}

where $z$ is the redshift, $D_{\text{L}}$ is the luminosity distance, $S_{\text{CO}}$ $\Delta v_{}$ is the velocity integrated flux in the radio definition, $\nu_{\text{obs}}$ is the observed frequency, and $\alpha_{\text{CO}}$ is the conversion factor. The \Ht~mass values are discussed in Sections \ref{im_2} -- \ref{im_4}.

\subsection{SALT Observations} \label{sec_salt}
A Southern African Large Telescope (SALT) long-slit observation tracing \textsc{[Oii]} along the major axis of galaxy 1432541 was conducted on 2023 May 24, in an attempt to detect an ionized gas rotation curve to confirm the internal kinematics of this galaxy [SALT 2022-2-DDT-003, PI: J. Blue Bird, CoPIs: M. Mogotsi, D.J. Pisano]. While H$\alpha$ is a primary indicator of star formation, \textsc{[Oii]} (indicating the presence of ionized oxygen in the gas of the galaxy) is a readily observable line that serves as a useful alternative, especially at redshifts beyond $z$ = 0.3 where H$\alpha$ becomes redshifted to the near infrared. Using the SFR and the H$\alpha$ / [OII] flux ratio \citep{Zhai2019}, we target a [OII] surface brightness sensitivity that requires an on-target exposure time of 4800 seconds. We use a grating that provides a resolution of R$\sim$4500 at the wavelength of the redshifted (5487 \textup{~\AA}) [OII] line (3727 \textup{~\AA}) and sampling at $\sim$ 15 km s$^{-1}$ with 4$\times$ binning. With the low signal-to-noise from the observation (peak SNR = 2.5), it is challenging to accurately measure the Doppler shift variations along the slit, which are essential for mapping the rotation curve. Although the observation was unable to produce a rotation curve, the peak of the \textsc{[Oii]} doublet was detected just above the noise at 5487\textup{~\AA}, and the center of the doublet is interpreted at $z = 0.473$. This coincides with the CO(1-0) line center observed with the LMT for this galaxy. The \textsc{[Oii]} spectrum is shown in Figure \ref{fig_salt}.

\subsection{JWST Observations} \label{sec_jwst}
We utilize high-resolution observations (0.06 arcsec / pixel) with JWST/NIRCam images from COSMOS-Web \citep{Casey2023} (and references therein), a contiguous 0.54 $\mathrm{deg^{2}}$ NIRCam imaging survey in four filters (F115W, F150W, F277W and F444W). These observations reach 5$\sigma$ point source depths ranging from $\sim$ 27.5 $-$ 28.2 mag. In parallel, the survey obtains 0.19 $\mathrm{deg^{2}}$ of MIRI imaging in one filter (F770W) that reaches 5$\sigma$ point source depths of $\sim$ 25.3 -- 26.0 magnitudes. The FOV of COSMOS-Web with NIRCam covers roughly two-thirds of the CHILES FOV, not covering areas to the east and north. Approximately half of this NIRCam area is covered by MIRI. Three out of the four galaxies presented in this study have NIRCam images, but none have usable MIRI imaging. The red, green and blue (RGB) images of the four NIRCam filters are constructed according to the Trilogy9 prescription \citep{Coe2012}. The filters are assigned to the red/green/blue channels as (R,G,B) = (F444W + F277W, F277W + F150W,  F150W + F115W), as seen in Figure \ref{fig_jwst}. 

\section{Results} 

\begin{figure}
\includegraphics[scale=0.68]{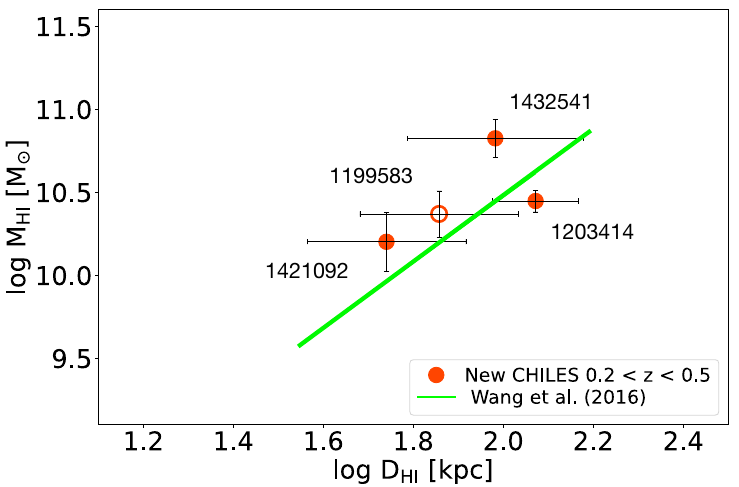}
\caption{
Relation of the \textsc{Hi} mass and \textsc{Hi} disk diameter. The radial extent of the \textsc{Hi} diameter D$_{HI}$ is measured along the \textsc{Hi} major axis of the PV diagram at a limiting column density of $1.25\times10^{20}$ cm$^{-2}$ (1 M$_{\odot}$ pc$^{-2}$). The open circle is the lowest stellar mass galaxy (M$_\mathrm{HI}$ $<$ $10^{10}$ M$_{\odot}$) in the sample and the solid circles are high stellar mass galaxies (M$_\mathrm{HI}$ $>$ $10^{10}$ M$_{\odot}$). The green line is the correlation found by \citet{Wang2016} for $z$ = 0. 
}
\label{fig_smr}
\end{figure}

At these large redshifts, we find that the galaxies host large \textsc{Hi} disks. As a sanity check, we see if our \textsc{Hi} disks fall on the \textsc{Hi} size--mass relation for local galaxies found by Wang et al. (2016), as shown in Figure \ref{fig_smr}. This size-mass relation is a useful verification of observed galaxy properties with both quantities measured directly from the data. For Galaxy 1432541, the 1$\sigma$ column density used is slightly above this level at $1.85\times10^{20}$ cm$^{-2}$. Errors in the \textsc{Hi} radial extent reflect uncertainties of one beam-width on either side. The \textsc{Hi} disks for the four galaxies are within 1$\sigma$ of the relation, indicating that the disks are not unusually large for these galaxies.

We study the spatially resolved distribution and morphology of the \textsc{Hi} gas compared to the optical extent. In the redshift range of our \textsc{Hi} detections ($z$ = 0.22 to 0.47), the synthesized beam ranges from 6.3$^{\prime\prime}$ to 7.5$^{\prime\prime}$, and the linear resolution ranges from 22 to 43 kpc, respectively. This translates to over two beam widths across the extent of our highest redshift galaxy.

\subsection{\textsc{Hi} Images} \label{sec_h1}
\begin{figure*}
\centering
\includegraphics[scale=1.1]{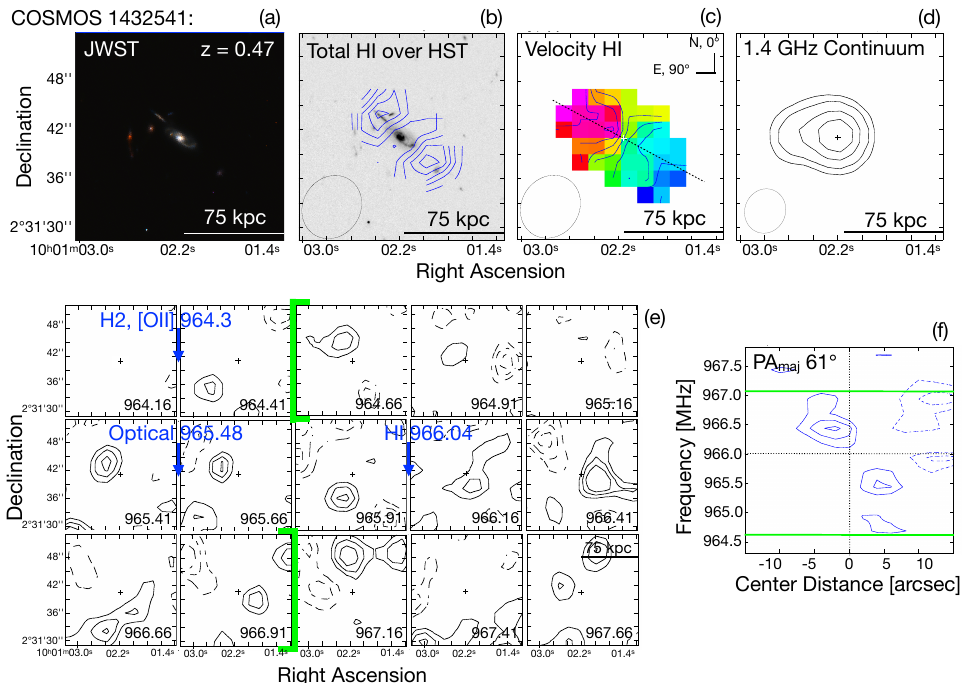}
\caption{
Galaxy 1432541 at $z$ = 0.47. \textbf{Panel a -- Near-IR JWST:} Four NIRCam filters combined (expanded image in Figure \ref{fig_jwst}). \textbf{Panel b -- Total \textsc{Hi} Intensity:} 6.46 (3.5$\sigma$), 8.62, 10.77, 12.93 $\times$ 10$^{20}$ cm$^{-2}$. \textbf{Panel c -- Intensity-Weighted \textsc{Hi} Velocity Field:} 966.04 MHz (system) $\pm$ 75 km s$^{-1}$, color range 964.53 -- 967.04 MHz. \textbf{Panel d -- 1.4 GHz Continuum:} 5, 10, 50, 150 $\mu \mathrm{Jy}$. \textbf{Panel e -- Channel Map:} RMS noise $\times$ $\pm$2, 3, and 4$\sigma$ (negative contours are dashed) Jy beam$^{-1}$, Channel width 78 km s$^{-1}$ (250 kHz), Total \textsc{Hi} integrated line width 772 km s$^{-1}$, Green brackets indicate the ten channels used to generate the total \textsc{Hi} intensity map and to calculate the \textsc{Hi} mass. \textbf{Panel f -- PV diagram:} RMS noise $\times$ $\pm$2, 3, and 4$\sigma$ (negative contours are dashed) Jy beam$^{-1}$. The northeast direction of the \textsc{Hi} velocity field (redshifted side) corresponds to positive distance from center. Detailed descriptions can be found in Sections \ref{sec_h1} and \ref{im_1}.
}
\label{fig_1432541im}
\end{figure*}

\begin{figure*}
\centering
\includegraphics[scale=1.1]{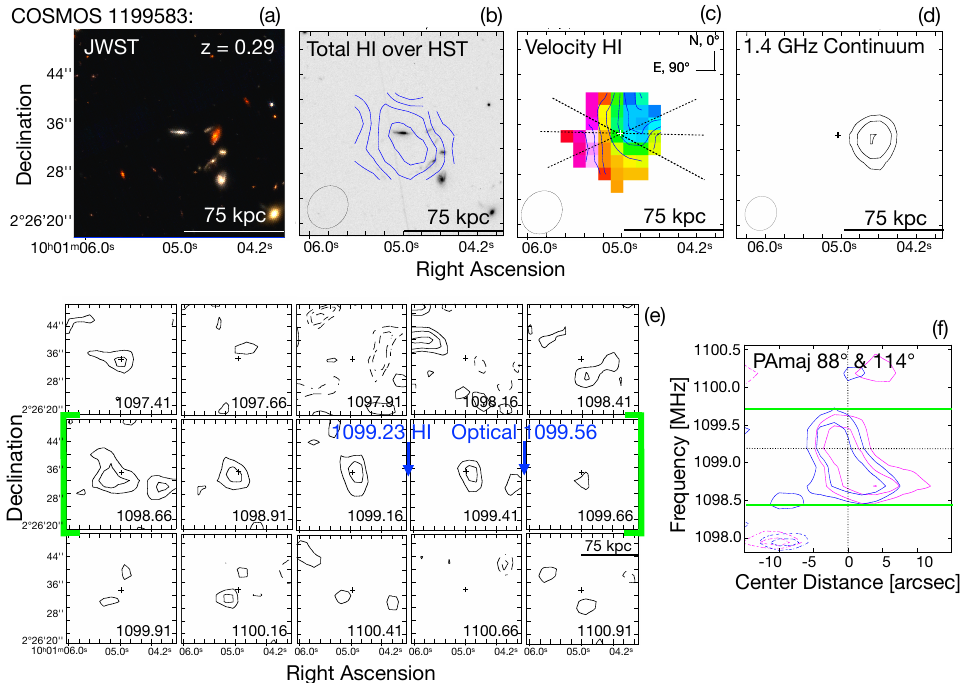}
\caption{
Galaxy 1199583 at $z$ = 0.29. \textbf{Panel a -- Near-IR JWST:} Four NIRCam filters combined (expanded image in Figure \ref{fig_jwst}). \textbf{Panel b -- Total \textsc{Hi} Intensity:} 1.85 (3$\sigma$), 4.31, 6.78, 9.24 $\times$ 10$^{20}$ cm$^{-2}$. \textbf{Panel c -- Intensity-Weighted \textsc{Hi} Velocity Field:} 1099.23 MHz (system) $\pm$ 35 km s$^{-1}$, color range 1098.53 -- 1099.79 MHz. \textbf{Panel d -- 1.4 GHz Continuum:} None coincident with the center; 5, 10, 50 $\mu \mathrm{Jy}$ at 26 kpc west of center. \textbf{Panel e -- \textsc{Hi} Channel Map:} RMS noise $\times$ $\pm$2, 3, and 4$\sigma$ (negative contours are dashed) Jy beam$^{-1}$, Channel width 68 km s$^{-1}$ (250 kHz), Total \textsc{Hi} integrated line width 334 km s$^{-1}$, Green brackets indicate the five channels used to generate the total \textsc{Hi} intensity map and to calculate the \textsc{Hi} mass. \textbf{Panel f -- PV diagram:} RMS noise $\times$ $\pm$2, 3, and 4$\sigma$ (negative contours are dashed) Jy beam$^{-1}$, Optical PA: 88$^{\circ}$ (blue contours), \textsc{Hi} PA:114$^{\circ}$ (magenta contours). The east direction of the \textsc{Hi} velocity field (redshifted side) corresponds to positive distance from center. Detailed descriptions can be found in Sections \ref{sec_h1} and \ref{im_2}.
}
\label{fig_1199583im}
\end{figure*}

\begin{figure*}
\centering
\includegraphics[scale=1.1]{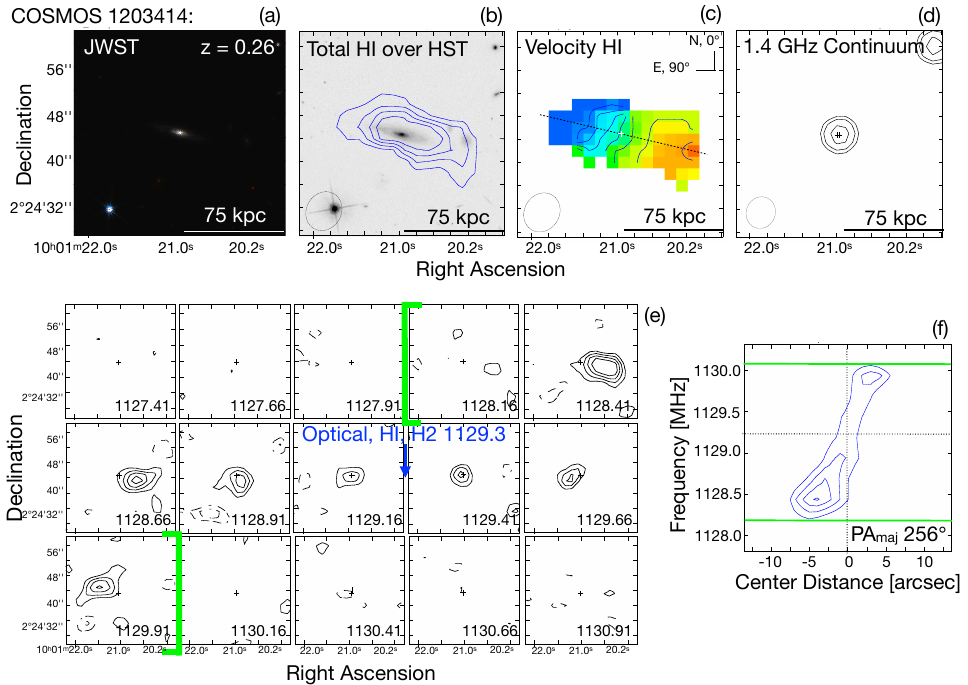}
\caption{
Galaxy 1203414 at $z$ = 0.26. \textbf{Panel a -- Near-IR JWST:} Four NIRCam filters combined (expanded image in Figure \ref{fig_jwst}). \textbf{Panel b -- Total \textsc{Hi} Intensity:} 2.81 (3$\sigma$), 5.62, 8.42, 11.23 $\times$ 10$^{20}$ cm$^{-2}$. \textbf{Panel c -- Intensity-Weighted \textsc{Hi} Velocity Field:} 1129.3 MHz (system) $\pm$ 65 km s$^{-1}$, color range 1128.13 -- 1130.04 MHz. \textbf{Panel d -- 1.4 GHz Continuum:} 5, 10, 20 $\mu \mathrm{Jy}$. \textbf{Panel e -- Channel Map:} RMS noise $\times$ $\pm$2, 3, and 4$\sigma$ (negative contours are dashed) Jy beam$^{-1}$, Channel width 66 km s$^{-1}$ (250 kHz), Total \textsc{Hi} integrated line width 527 km s$^{-1}$, Green brackets indicate the eight channels used to generate the total \textsc{Hi} intensity map and to calculate the \textsc{Hi} mass. \textbf{Panel f -- PV diagram:} RMS noise $\times$ $\pm$2, 3, and 4$\sigma$ (negative contours are dashed) Jy beam$^{-1}$, The east direction of the \textsc{Hi} velocity field (blue-shifted side) corresponds to positive distance from center. Detailed descriptions can be found in Sections \ref{sec_h1} and \ref{im_3}.
}
\label{fig_1203414im}
\end{figure*}

\begin{figure*}
\centering
\includegraphics[scale=1.1]{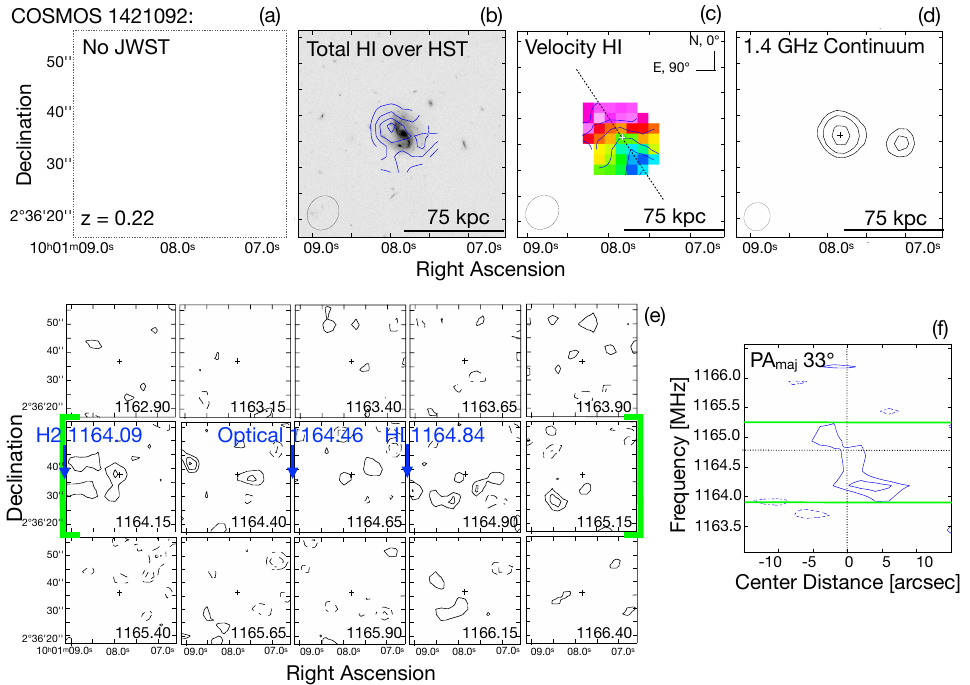}
\caption{
Galaxy 1421092 at $z$ = 0.22. \textbf{Panel a -- Near-IR JWST:} No image for this galaxy. \textbf{Panel b -- Total \textsc{Hi} Intensity:} 2.03 (3$\sigma$), 4.05, 6.08, 8.10 $\times$ 10$^{20}$ cm$^{-2}$. \textbf{Panel c -- Intensity-Weighted \textsc{Hi} Velocity Field:} 1164.85 MHz (system) $\pm$ 65 km s$^{-1}$, color range 1164.02 -- 1165.28 MHz. \textbf{Panel d -- 1.4 GHz Continuum:} 5, 10, 35 $\mu \mathrm{Jy}$. \textbf{Panel e -- Channel Map:} RMS noise $\times$ $\pm$2, 3, and 4$\sigma$ (negative contours are dashed) Jy beam$^{-1}$, Channel width 64 km s$^{-1}$ (250 kHz), Total \textsc{Hi} integrated line width 316 km s$^{-1}$, Green brackets indicate the five channels used to generate the total \textsc{Hi} intensity map and to calculate the \textsc{Hi} mass. \textbf{Panel f -- PV diagram:} RMS noise $\times$ $\pm$1.75, 2.50$\sigma$ (negative contours are dashed) Jy beam$^{-1}$. The northeast direction of the \textsc{Hi} velocity field (redshifted side) corresponds to positive distance from center. Detailed descriptions can be found in Sections \ref{sec_h1} and \ref{im_4}.
}
\label{fig_1421092im}
\end{figure*}

We present the images for individual galaxies in Figures \ref{fig_1432541im} -- \ref{fig_1421092im}. All images and channel maps show a region of 150 kpc$\times150$ kpc centered on the galaxy. All nearby background galaxies with known redshifts in the JWST images are confirmed to be distant enough to be ruled out of association or interaction. Other smaller galaxies seen in the JWST and HST images are not close in redshift to our target galaxies and are a mix of foreground and background sources. The targets do not have companions within 200 kpc and $z$ $\pm$ 0.006. We will discuss possible spectroscopic group membership (one galaxy has a photometric neighbor with a lower membership probability) for each galaxy separately using the group catalog of \citet{Knobel2012}, analogous to \citet{Hess2019}.

The total \textsc{Hi} intensity or moment 0 maps of each galaxy are overlaid as contours on the HST images. The VLA synthesized beam is shown in the lower left corner. The \textsc{Hi} intensity-weighted velocity fields or moment 1 maps are velocity contours overlaid on their own color scale. The optical center is marked with a white cross. The optical and \textsc{Hi} major axes are shown as dotted lines with the position angle (PA) of the receding side. In the velocity field, the line passing through the white cross represents the \textsc{Hi} systemic velocity. The 1.4 GHz continuum emission images and intensities are from the CHILES Con Pol survey \citep{Luber2025b, Gim2025}. 

In lieu of the \textsc{Hi} spectrum, we include the \textsc{Hi} channel maps. The lowest contours of the \textsc{Hi} channel maps represent $\pm$2$\sigma$, where $\sigma$ is the RMS per beam per channel. The channel width is 250 kHz.  In every channel map, the optical center is shown with a black cross, and the frequency (MHz) is shown in the lower left corner. The green brackets in the \textsc{Hi} channel maps indicate the channels used to generate the \textsc{Hi} intensity map and calculate the \textsc{Hi} mass. The frequency relating to the green brackets is shown in the PV diagram in green. The horizontal dotted line is the \textsc{Hi} systemic velocity, and the PA used to make the two-dimensional position-velocity slice is also shown in the PV diagram. 

The individual parameter values and contour levels are listed in the captions of Figures \ref{fig_1432541im} -- \ref{fig_1421092im}. 

\subsubsection{Galaxy 1432541} \label{im_1} 
Figure \ref{fig_1432541im} shows massive spiral galaxy 1432541 (Panel a) at an optical redshift of $z$ = 0.4712. It has a stellar mass of $1.9\times10^{11}$ M$_{\odot}$, an \textsc{Hi} mass of $6.7 \pm 1.7\times10^{10}$ M$_{\odot}$ and an \Ht~mass of $5.2 \pm 1.6\times10^{10}$ M$_{\odot}$ (Table \ref{tab_prop}). The SFR is 29.9 M$_{\odot}$ yr$^{-1}$ (SFR$_\mathrm{1.4}$ = 36.5 M$_{\odot}$ yr$^{-1}$) and there is 1.4 GHz continuum emission of 307 $\mu\mathrm{Jy}$ with radio luminosity of 2.1$\times10^{23}$ W Hz$^{-1}$ (Panel d), with star-forming galaxies dominating in the range of $\sim$ $10^{22} - 10^{23}$ W Hz$^{-1}$ \citep{radcliffe2021}. The near-IR JWST image (Figure \ref{fig_1432541im} -- Panel a and Figure \ref{fig_jwst} -- Panel a) shows a disturbance in the stellar disk, with lopsided spiral arms and a distinct knot about 8 kpc from the center. The image suggests that there may have been past encounters with other galaxies, which could be a reason for the high SFR. The galaxy belongs to a two-member group in the Knobel catalog \citep{Knobel2012} and has a nearby spectroscopic companion separated by 260 kpc to the southeast and by 210 km s$^{-1}$ at $z$ = 0.4719. The companion has a stellar mass of $1.0\times10^{10}$ M$_{\odot}$ and is not detected in \textsc{Hi} in our data. 

This galaxy has an \textsc{Hi} extent of 96 $\pm$ 43 kpc, 1.7 $\pm$ 0.8 times the optical extent of 57 kpc. The \textsc{Hi} morphology is fairly symmetric with a slight asymmetry to the northwest in the total \textsc{Hi} intensity map (Panel b), with falling \textsc{Hi} contours towards the center. The galaxy is inclined at 67$^{\circ}$. The \textsc{Hi} position angle is aligned with the optical position angle, and the \textsc{Hi} velocity field (Panel c) is symmetric along the position angles. The contour running through the white cross in the center of the galaxy (Panel c) corresponds to the \textsc{Hi} systemic velocity at a frequency of 966.04 MHz. The channel map (Panel e) shows, in blue text, this central frequency of the \textsc{Hi} plus \textsc{Hi}-rest-frequency equivalents for the optical, \textsc{[Oii]} and derived \Ht. The \textsc{Hi} mass and the \textsc{Hi} intensity map (Panel b) are produced using the ten channels between the green brackets. There is a partial stripe in the lower left channel at 966.66 MHz over the emission; therefore, only half of the \textsc{Hi} flux in that channel is used in the calculation of the \textsc{Hi} mass, a similar value to the \textsc{Hi} flux in the following channel at 966.91 MHz. The emission in these lower frequencies corresponds to the side of the sky where the compact knot is in the JWST image (Figure \ref{fig_jwst} -- Panel a).

The observed velocity difference between the \textsc{Hi} and optical with the \Ht~and \textsc{[Oii]} indicates complex gas kinematics. We emphasize here that the \textsc{Hi} systemic velocity seems well determined based on the \textsc{Hi} velocity field. Such offsets in velocity are often seen in active galactic nuclei (AGN) or regions of intense star formation. This observed velocity difference may suggest bulk outflow motion of molecular and ionized gas driven by star formation processes. With $\sim$ 160,000 star-forming SDSS galaxies at a median $z$ = 0.073, \citet{Cicone2016} find a tight correlation of increased ionized outflows with increasing SFR for those located above the star-forming main sequence (SFMS) and $>$ 1 M$_{\odot}$ yr$^{-1}$. Moreover, with a sample of 13 AGN and star-forming galaxies (M$_{*}$ = $10^{10.4-11.6}$ M$_{\odot}$) at $z$ $<$ 0.2, \citet{Fluetsch2019} find the ionized outflow is about as massive as the molecular outflow in starburst galaxies. Galaxy 1432541 is also a starburst galaxy and is likely to have an outflow. The optical images suggest an ejection of gas on the northeast side (the redshifted side). However, there are some differences from the results quoted above. The \textsc{[Oii]} lines are redshifted, whereas statistically the optical outflows appear blueshifted as the far side of the galaxy gets obscured by dust. In addition, an outflow is usually identified by a second broad component in the profile. In our case, the molecular profile has only one very redshifted component, which is slightly narrower than the \textsc{Hi}. This may suggest that there is no molecular gas associated with the galaxy itself.  An alternative interpretation is that the galaxy was tidally disturbed and that we are seeing the fallback of material from the northeast side. This material would be on the near side of the galaxy, and it would explain the (unusually highly) redshifted velocities and possibly some \textsc{Hi} absorption. It is worth noting that the low SNR of all measurements may play a role. Additional observations, including images of the molecular and ionized gas, may provide clues to understanding the complex gas kinematics of this system.

\subsubsection{Galaxy 1199583} \label{im_2}
Figure \ref{fig_1199583im} shows galaxy 1199583 (Panel a) at an optical redshift of $z$ = 0.2918. The \textsc{Hi} mass is $2.3 \pm 0.8\times10^{10}$ M$_{\odot}$, which is 13 times the dwarf-like stellar mass of $1.7\times10^{9}$ M$_{\odot}$. CO is not detected in this galaxy and the \Ht~mass upper limit is $2.7\times10^{9}$ M$_{\odot}$ (Table \ref{tab_prop}). The SFR is 3.9 M$_{\odot}$ yr$^{-1}$ and there is no 1.4 GHz continuum emission coincident with the center of the galaxy (Panel d). The near-IR JWST image (Figure \ref{fig_1199583im} -- Panel a and Figure \ref{fig_jwst} -- Panel b) shows asymmetry, along with a distinct knot, on the east side. In the individual HST and JWST filters, the overall morphology does not change much but appears brighter at bluer wavelengths. This galaxy has a high star formation efficiency (SFE = SFR / M$_{HI}$ = 0.18 Gyr$^{-1}$). It is not part of a group according to the Knobel catalog \citep{Knobel2012}. The nearest neighbor is located over 1 Mpc to the east, separated by 162 km s$^{-1}$ at $z$ = 0.2917, and is not detected in \textsc{Hi} in our data.

The galaxy has an \textsc{Hi} extent of 72 $\pm$ 29 kpc, which is 3.8 $\pm$ 1.5 times the optical extent of 19 kpc. The \textsc{Hi} morphology (Panel b) is symmetric around the optical center of the galaxy with extended emission to the south, and the \textsc{Hi} PA (Panel c) is offset from the optical PA by 26$^{\circ}$. The galaxy is inclined at 75$^{\circ}$. The \textsc{Hi} systemic velocity at a frequency of 1099.23 MHz differs from the optical redshift by -92 km s$^{-1}$. The channel map (Panel e) shows this central frequency of the \textsc{Hi} and the \textsc{Hi}-rest-frequency equivalent of the optical redshift. The \textsc{Hi} mass and \textsc{Hi} intensity map (Panel b) are produced using the range of five channels in-between the green brackets, which corresponds to an \textsc{Hi} integrated line width of 334 $\pm$ 68 km s$^{-1}$. The PV diagram (Panel f) shows a slice along both the optical PA and \textsc{Hi} PA.

\subsubsection{Galaxy 1203414} \label{im_3}
Figure \ref{fig_1203414im} shows spiral galaxy 1203414 (Panel a) at an optical redshift of $z$ = 0.2577. It has a stellar mass of $7.1\times10^{10}$ M$_{\odot}$, an \textsc{Hi} mass of $2.8 \pm 0.4\times10^{10}$ M$_{\odot}$ and an \Ht~mass of $4.4 \pm 2.7\times10^{9}$ M$_{\odot}$ (Table \ref{tab_prop}). The SFR is 2.5 M$_{\odot}$ yr$^{-1}$ (SFR$_\mathrm{1.4}$ = 6.6 M$_{\odot}$ yr$^{-1}$), and there is 1.4 GHz continuum emission of 43 $\mu\mathrm{Jy}$ with radio luminosity of 9.0$\times10^{21}$ W Hz$^{-1}$ (Panel d). The near-IR JWST image (Figure \ref{fig_1203414im} -- Panel a and Figure \ref{fig_jwst} -- Panel c) shows that this galaxy has an undisturbed stellar component, which is interestingly contrary to the asymmetric \textsc{Hi}. This galaxy belongs to a two-member group in the Knobel catalog \citep{Knobel2012} and has a photometric neighbor (lower membership probability) that is not detected in \textsc{Hi} in our data, separated by 250 kpc to the northeast and by 216 km s$^{-1}$ at $z$ = 0.2581.

The galaxy has an \textsc{Hi} diameter of 118 $\pm$ 26 kpc, 2.3 $\pm$ 0.5 times the optical extent of 51 kpc. The \textsc{Hi} morphology (Panel b) is extended to the northeast and northwest, plus very extended asymmetric morphology to the southwest. The velocity field (Panel c) is also asymmetric in the northeast and the southwest. The \textsc{Hi} systemic velocity at a frequency of 1129.3 MHz aligns with the \textsc{Hi} rest-frequency equivalents for the optical redshift and the derived \Ht, as shown in the channel map (Panel e). The \textsc{Hi} mass and \textsc{Hi} intensity map (Panel b) are produced using the eight channels between the green brackets, which corresponds to an \textsc{Hi} integrated line width of 527 $\pm$ 66 km s$^{-1}$. The galaxy is inclined at 45$^{\circ}$. The \textsc{Hi} position angle is aligned with the optical position angle, and the PV diagram (Panel f) shows more \textsc{Hi} on the lower frequency (blue-shifted) side.

\subsubsection{Galaxy 1421092} \label{im_4}
Figure \ref{fig_1421092im} shows spiral galaxy 1421092 (Panel a) at an optical redshift of $z$ = 0.2198. It has a stellar mass of $2.7\times10^{10}$ M$_{\odot}$, an \textsc{Hi} mass of $1.6 \pm 0.7\times10^{10}$ M$_{\odot}$ and an \Ht~mass of $6.3 \pm 5.5\times10^{9}$ M$_{\odot}$ (Table \ref{tab_prop}). The SFR is 5.8 M$_{\odot}$ yr$^{-1}$ (SFR$_\mathrm{1.4}$ = 13.1 M$_{\odot}$ yr$^{-1}$) and there is 1.4 GHz continuum emission of 130 $\mu\mathrm{Jy}$ with radio luminosity of 1.8$\times10^{22}$ W Hz$^{-1}$ (Panel d). There is no JWST image for this galaxy. The i-band HST image (Panel b) shows that this galaxy has strong spiral-arm features, with a visible clump to the southwest. This galaxy belongs to a group with more than a dozen members in the Knobel catalog \citep{Knobel2012} and has a spectroscopic neighbor, not detected in \textsc{Hi} in our data, separated by 383 kpc to the southeast and by 30 km s$^{-1}$ at $z$ = 0.2197.

The galaxy has an \textsc{Hi} extent of 55 $\pm$ 22 kpc, 1.3 $\pm$ 0.5 times the optical extent of 41 kpc. The \textsc{Hi} morphology (Panel b) is very asymmetric with more emission to the northeast, and the \textsc{Hi} kinematics (Panel c) are warped to the northeast as well. The channel map (Panel e) shows the \textsc{Hi} systemic velocity at a frequency of 1164.85 MHz, which differs from the \textsc{Hi} rest-frequency equivalents for the optical redshift by +98 km s$^{-1}$ and the derived \Ht~by +194 km s$^{-1}$. The \textsc{Hi} mass and \textsc{Hi} intensity map (Panel b) are produced using the five channels between the green brackets, which corresponds to an \textsc{Hi} integrated line width of 316 $\pm$ 64 km s$^{-1}$. The galaxy is inclined at 79$^{\circ}$. The \textsc{Hi} PA aligns with the optical PA and the PV diagram (Panel f) shows more \textsc{Hi} on the lower frequency (redshifted) side.	

\subsubsection{Summary of Galaxy Features}
Our sample of four galaxies consists of three high stellar mass galaxies (M$_{*}$ $>$ $10^{10}$ M$_{\odot}$ ) and one low stellar mass galaxy (M$_{*}$ $<$ $10^{10}$ M$_{\odot}$). See Table \ref{tab_gal} for a summary of features. Section \ref{sec_h1h2} will further analyze the galaxies grouped in these two mass bins. All four galaxies have extended \textsc{Hi} (1.3 -- 3.8$\times$ optical extent), high \textsc{Hi} mass ($1.4 - 6.7\times10^{10}$ M$_{\odot}$ ), and high SFR (3 -- 30 M$_{\odot}$ yr$^{-1}$). All three high stellar mass galaxies are detected in \Ht~($0.4 - 5.2\times10^{10}$ M$_{\odot}$ ), have 1.4 GHz continuum emission (43 -- 307 $\mu\mathrm{Jy}$) and are part of small groups, each with nearby neighbors (250 -- 383 kpc in the sky), undetected in \textsc{Hi}. Interestingly, all four galaxies have some asymmetry in the \textsc{Hi} but only two (both with high stellar mass) have disturbed stellar components.

\subsection{Cold Gas Properties} \label{sec_h1h2}
We explore how our sample of four higher-redshift galaxies matches samples at other redshifts. We combine these into a total CHILES sample of 19 galaxies that have measurements in both \textsc{Hi} and \Ht~spanning 0.1 $<$ $z$ $<$ 0.5 \citep{Fernandez2016, Hess2025}. We compare these galaxies to \textsc{Hi} surveys that have CO follow-up observations: from the local Universe, xCOLD GASS \citep{Saintonge2011, Saintonge2017}, CO in HIghMass \citep{Hallenbeck2016}, and CO in \textsc{Hi} Monsters \citep{Lee2014}; plus redshift $z$ $\sim$ 0.2 galaxies, CO in HIGHz \citep{Cortese2017}, and COOL BUDHIES \citep{Cybulski2016} (and references therein). This provides, for the first time, a continuous look at directly detected \textsc{Hi} and \Ht~in emission in individual galaxies over the redshift range 0 $<$ $z$ $<$ 0.5. This work builds on the analysis in \citet{Hess2025} by essentially doubling the look-back time and focusing on high stellar mass (M$_\mathrm{*}$ $>$ $10^{10}$ M$_{\odot}$) and high \textsc{Hi} mass (M$_\mathrm{HI}$ $>$ $10^{10.5}$ M$_{\odot}$) samples. We discuss impacting factors to the analysis such as the CO conversion factor, sample selection, and \textsc{Hi} flux-limitations. The surveys are summarized in Table \ref{tab_samples}.

\begin{table*}	
\caption{Galaxy Features}			
\begin{adjustwidth}{-1.5cm}{0cm}			
\centering			
\begin{tabular}{ccc|c}			
\hline			
\hline			
Galaxy 1432541&	Galaxy 1203414&	Galaxy 1421092&	Galaxy 1199583\\
$z$ = 0.47&	$z$ = 0.26&	$z$ = 0.22&	$z$ = 0.29\\[1.7ex]
\hline			
High stellar mass*&	High stellar mass&	High stellar mass&	Low stellar mass\\
High \textsc{Hi} mass, \textsc{Hi} 1.7x optical&	High \textsc{Hi} mass, \textsc{Hi} 2.3x optical&	High \textsc{Hi} mass, \textsc{Hi} 1.3x optical& 	High \textsc{Hi} mass, \textsc{Hi} 3.8x optical\\ 
\Ht~detected&	\Ht~detected&	\Ht~detected&	\Ht~upper limit\\
\Ht, Opt lines redshifted from \textsc{Hi}&	\Ht, Opt lines align with \textsc{Hi}&	\Ht, Opt lines redshifted from \textsc{Hi}& 	Opt line blue-shifted from \textsc{Hi}\\[1.7ex]
\hline			
High SFR&	High SFR&	High SFR&	High SFR\\
SFMS $z$ $\sim$ 0.8&	SFMS $z$ $\sim$ 0.15&	SFMS $z$ $\sim$ 0.4&	SFMS $z$ $\sim$ 1.7\\
Continuum emission&	Continuum emission&	Continuum emission&	No continuum\\
Disturbed stellar in JWST&	Undisturbed stellar in JWST&	Undisturbed stellar in HST&	Disturbed stellar in JWST\\
Nearby neighbor&	Nearby neighbor&	Nearby neighbor&	Not in group\\[1.7ex]
\hline			
\hline			
\multicolumn{4}{p{16.85cm}}{\footnotesize{The vertical line separates the three galaxies with M$_\mathrm{*}$ $>$ $10^{10}$ M$_{\odot}$ from the one galaxy with M$_\mathrm{*}$ $<$ $10^{10}$ M$_{\odot}$. *M$_\mathrm{*}$ $>$ $10^{11}$ M$_{\odot}$.}}
\end{tabular}			
\label{tab_gal}			
\end{adjustwidth}			
\end{table*}

The total CHILES \textsc{Hi} + \Ht~sample is comprised of 14 (at $z$ = 0.12) of 16 galaxies from \citet{Hess2019}, of which five have direct \Ht~measurements from ALMA (CO(1--0)), nine have \Ht~upper limits, together with one stacked \Ht~detection \citep{Hess2025}. We include the galaxy at $z$ = 0.38 from \cite{Fernandez2016} with \Ht~measurement from LMT (CO(1--0)). ALMA CO (3-2) observations exist for this galaxy and are presented in an upcoming analysis (Donovan Meyer et al. in preparation). In addition, our four new \textsc{Hi} detections have three direct \Ht~measurements from the LMT (CO(1--0)) and one \Ht~upper limit. We also incorporate six stacked CHILES \textsc{Hi} detections of blue star-forming galaxies (NUV-r $<$ 3), in two stellar mass bins of $10^{10-12.5}$ M$_{\odot}$ and $10^{9-10}$ M$_{\odot}$ at three redshift ranges ($z$ = 0.15, 0.31, and 0.41) from \citet{Luber2025a} (and references therein).

We compare our combined CHILES sample to blue galaxies (NUV-r $<$ 3) with \Ht~measurements from the xCOLD GASS survey at $z$ $<$ 0.05, consisting of 82 galaxies with M$_{*}$ $<$ $10^{10}$ M$_{\odot}$ and 65 galaxies with M$_{*}$ $>$ $10^{10}$ M$_{\odot}$. We add follow-up \Ht~measurements of three galaxies from HIghMass at $z$ $<$ 0.06. We include \Ht~measurements in 15 \textsc{Hi} Monster galaxies with \textsc{Hi} mass $>$ $10^{10.5}$ M$_{\odot}$ at $z$= 0.04 -- 0.05 \citep{Lee2014}. We add follow-up \Ht~measurements of five blue galaxies (NUV-r $<$ 3) from HIGHz at $z$ $\sim$ 0.2. In addition, we include the median values from 9153 \textsc{Hi}-selected ALFALFA galaxies with SDSS spectra at $z$ $<$ 0.06 \citep{Maddox2015}. Furthermore, we include six galaxies with \textsc{Hi} plus \Ht~measurements and six galaxies with \textsc{Hi} plus \Ht~upper limits from COOL BUDHIES at $z$ $\sim$ 0.2.

Two adjustments have been made to the data in this comparison. First, the Jy/K gain factor for the RSR on the LMT was re-derived in 2020 using a variety of new and archival calibration measurements, with existing data supporting a frequency-dependent gain. The \Ht~values derived from LMT CO(1--0) measurements in COOL BUDHIES and for \citet{Fernandez2016} have been reduced with a frequency-dependent correction (RSR Gain Factor Calibration, 2020, M. Yun) of 13-14\% and 18\%, respectively. We also apply this frequency-dependent correction in our own LMT data analysis. Second, the \Ht~masses for the CHILES samples are calculated assuming an $\alpha_{\text{CO}}$ = 4.3 (includes the contribution of helium), the galactic CO-to-\Ht~conversion factor $\alpha_{\text{CO}}$ that is typically assumed for the Milky Way \citep{Bolatto2013, Saintonge2022}. The HIghMass, \textsc{Hi} Monsters, and HIGHz measurements use a CO conversion factor of 3.2 (does not include the contribution of helium). Their \Ht~values are adjusted to reflect $\alpha_{\text{CO}}$ = 4.3 in the plots, which is consistent with the average of the xCOLD GASS sample and previous CHILES samples. A continued discussion of the conversion factor follows in Section \ref{convf}.

\begin{figure*}
\centering
\includegraphics[scale=0.87]{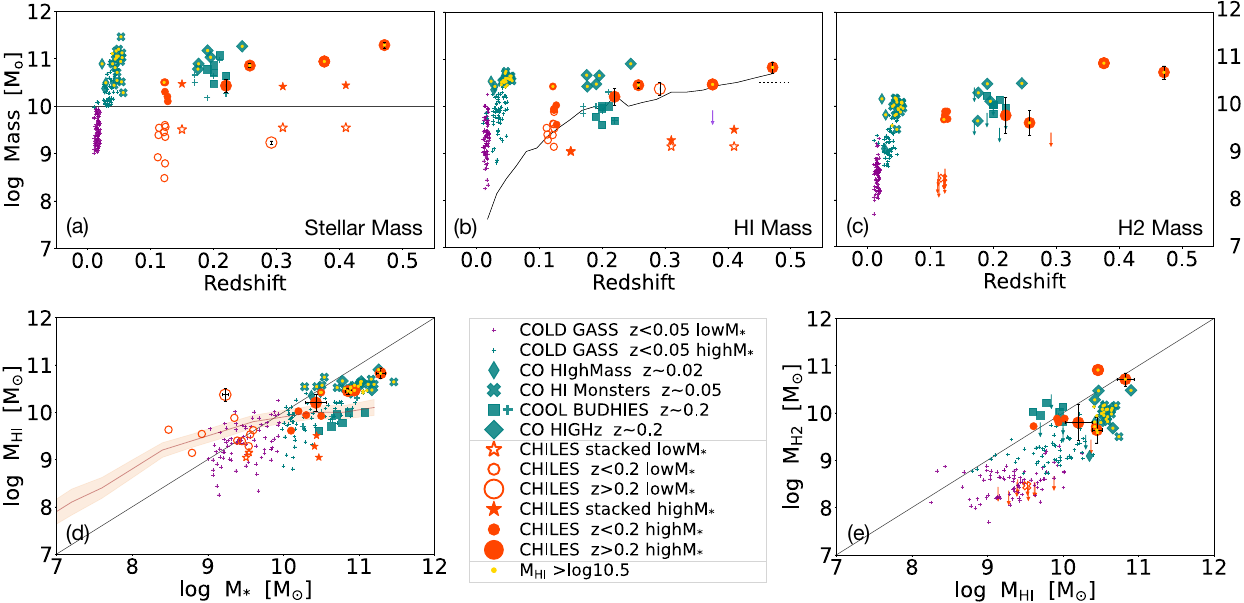}
\caption{
Mass properties of galaxies. The samples are divided into two stellar mass bins. Above $10^{10}$ M$_{\odot}$ have filled orange symbols and teal symbols. Below $10^{10}$ M$_{\odot}$ have open orange symbols and purple symbols. Galaxies that have \textsc{Hi} mass $>$ $10^{10.5}$ M$_{\odot}$ are overlaid with small gold circles. The purple and small teal crosses are the xCOLD GASS sample at $z$ $<$ 0.05. The teal slim diamonds are from HIghMass with CO at $z$ = 0.02 -- 0.03. The teal Xs are \textsc{Hi} Monsters with CO at $z$ = 0.04 -- 0.05. The teal squares are from HIGHz with CO at $z$ = 0.17 -- 0.25. The teal diamonds are from the COOL BUDHIES survey at $z$ = 0.16 -- 0.22 and the thick teal crosses are COOL BUDHIES galaxies with \Ht~upper limits. The CHILES sample is divided into two redshift ranges, $z$ $<$ 0.2 (small orange circles) and $z$ $>$ 0.2 (large orange circles). The orange stars are CHILES stacked results at $z$ = 0.15, 0.31, and 0.41. The open orange X is a stacked \Ht~detection at $z$ $\sim$ 0.12 from \citet{Hess2025}. \textbf{Panel a:} Stellar mass versus redshift. \textbf{Panel b:} \textsc{Hi} mass versus redshift. The solid black line is the \textsc{Hi} mass detection limit assuming 200 km s$^{-1}$ at 5$\sigma$ in the center of the field of view. The two adjacent orange circles at $z$ = 0.47 show the \textsc{Hi} mass calculated using both five channels and ten channels. The light purple arrow is an \textsc{Hi} upper limit from \citet{Heywood2024}. \textbf{Panel c:} \Ht~mass versus redshift. \textbf{Panel d:} \textsc{Hi} mass versus stellar mass. The brown line and shaded region shows the median values from the sample in \citet{Maddox2015} at $z$ $<$ 0.06. The black diagonal line represents equal parts stars and gas. \textbf{Panel e:} \textsc{Hi} mass versus \Ht~mass. The black diagonal line represents equal parts \textsc{Hi} and \Ht~gas.
}
\label{fig_mass}
\end{figure*}

Since so few measurements of both \textsc{Hi} and \Ht~exist beyond the local Universe, we compare all the existing samples together. In order to understand the sample selection, their properties are summarized in Table \ref{tab_samples}. The samples are flux-limited, have varied selection criteria, and the galaxies lie in various environments. At $z$ $\sim$ 0.2, COOL BUDHIES galaxies are in denser environments, and HIGHz galaxies are in more isolated environments but, when combined, may represent a fuller range of typical galaxies at their redshift. The methods for calculating the stellar mass and SFR are listed, with most samples using UV to IR bands. A NUV-r $<$ 3 cut is applied when measurements are available. As mentioned above, the samples are adjusted to use a similar CO conversion factor. Moreover, we are comparing global \Ht~masses which are typically best returned with single-dish telescopes so as to not filter out spatial information. The CO(1--0) measurements in this analysis are made mostly with single-dish telescopes (Table \ref{tab_samples}). The exceptions are the ALMA measurements for CHILES galaxies at $z \sim$ 0.12 \citep{Hess2025} and HIGHz galaxies with CO \citep{Cortese2017}, both of which point out in their analyses that they are not missing a significant fraction of the total flux.

While there are substantial selection effects with much of the higher-redshift data, biased towards star-forming gas-rich objects, the statistical trends are to be confirmed with more unbiased samples which are representative of the general trend of normal galaxies which contain the bulk of the baryonic mass.

\subsubsection{Cold Gas Mass}  \label{sec_mass}
We examine the mass content of our samples as a function of redshift, shown in Figure \ref{fig_mass}. Panel a shows the stellar mass versus redshift. Galaxies at different stellar masses evolve differently; we therefore divide the samples into two stellar mass bins, below and above $10^{10}$ M$_{\odot}$, separated by the black line. There are 102 galaxies above $10^{10}$ M$_{\odot}$ including 20 beyond the local Universe ($z$ $>$ 0.06). Indicated on the plots (overlaid gold circles) are 29 galaxies that have \textsc{Hi} mass $>$ $10^{10.5}$ M$_{\odot}$, including nine beyond the local Universe ($z$ $>$ 0.06). The CHILES sample consists of ten galaxies below $10^{10}$ M$_{\odot}$ and nine above.

The \textsc{Hi} mass versus redshift is shown in Panel b. The solid black line is the CHILES \textsc{Hi} mass detection limit assuming 200 km s$^{-1}$ at 5$\sigma$ in the center of the field of view; CHILES is detecting \textsc{Hi} in galaxies just above our detection limit across the redshift range. The short dotted line on the top right represents the survey goal of being able to detect $3\times10^{10}$ M$_{\odot}$ at our highest redshift. Panel c shows \Ht~mass versus redshift. Galaxies at higher redshift have more \Ht~than their local Universe counterparts with similar stellar mass, at least those that have \textsc{Hi} content above our detection limits. 

The \textsc{Hi} mass versus stellar mass is shown in Panel d. The median values obtained from ALFALFA \citep{Maddox2015} in the local Universe are shown as a brown line with a 1$\sigma$ shaded band. The CHILES sample mostly follow the band of median values for M$_{*}$ $<$ $10^{10.5}$ M$_{\odot}$. As expected, the stacked values of CHILES are well below the median. The high mass CHILES sample falls below the black line, indicating that these galaxies contain more stars than \textsc{Hi}. Panel e shows the \Ht~mass versus \textsc{Hi} mass. The high mass CHILES sample falls closer to the black line of equal amounts of \textsc{Hi} and \Ht, more than the local Universe samples.

We include the 3$\sigma$ upper limit (Panel b, purple arrow) for the CHILES galaxy at $z=0.38$ \citep{Fernandez2016} obtained by \citet{Heywood2024} in the MIGHTEE-HI survey. The upper limit of $8.1\times10^{9}$ M$_{\odot}$ is a factor of three lower than the CHILES measurement of $2.9 \pm 1.0\times10^{10}$ M$_{\odot}$. Both measurements are affected by RFI, which has increasingly become a problem between 2013, when epoch 1 of CHILES was observed, and today. As a result, we have not been able to confirm as yet the result found in Epoch 1. Although the CHILES result is consistent with scaling relations to within one sigma, if the MIGHTEE-HI measurement is accurate, then this galaxy would be extremely \textsc{Hi}-poor given its other known properties. Other MIGHTEE-HI detections \citep{Jarvis2025}, shown in Section \ref{beyond}, have \textsc{Hi} masses that are at least three times as large as their upper limit for the CHILES galaxy at $z$ = 0.38, casting some doubt on the validity of the limit. Either way, this is a very interesting galaxy.

\begin{figure}
\centering
\includegraphics[scale=1.53]{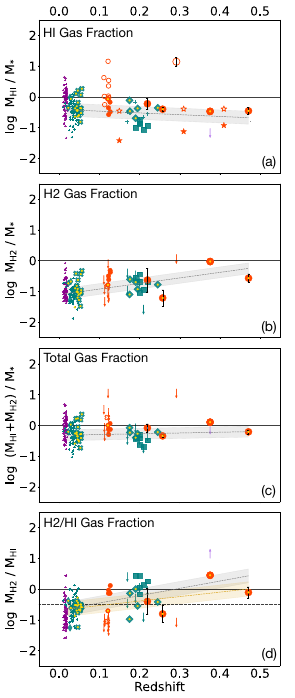} 
\caption{
Gas fraction properties of galaxies. Gas fraction properties of galaxies. The samples as well as the legend are the same as in Figure \ref{fig_mass}. The dashed silver lines are the best fit to 102 high stellar mass galaxies (M$_{*}$ $>$ $10^{10}$ M$_{\odot}$) that have both \textsc{Hi} and \Ht~detections (solid orange and teal symbols, not including six COOL BUDHIES galaxies with \Ht~upper limits (thick teal crosses) and not including stacked masses). The dashed gold line is the best fit for this high stellar mass sample with for 29 high \textsc{Hi} mass galaxies with M$_\mathrm{HI}$ $>$ $10^{10.5}$ M$_{\odot}$ (small gold circles overlaid). Both lines have 1$\sigma$ shaded areas. The light purple arrow is an \textsc{Hi} upper limit from \citet{Heywood2024}. \textbf{Panel a:} \textsc{Hi} gas fraction versus redshift. \textbf{Panel b:} \Ht~gas fraction versus redshift. \textbf{Panel c:} Total fraction versus redshift. \textbf{Panel d:} \Ht~/ \textsc{Hi} gas fraction versus redshift.
}
\label{fig_gf}
\centering
\end{figure}

\subsubsection{Cold Gas Fraction}
Using a combination of simulations, observations and theoretical models to understand how \textsc{Hi} and \Ht~gas evolve across cosmic time, results from \citet{Walter2020} show that the \textsc{Hi} cosmic density remains approximately constant with redshift, with a decline by a factor of two from the peak of cosmic star formation to today, with \Ht~cosmic density decreasing significantly, by a factor of six.  \citet{Hess2025} find mild evidence of evolution in the \Ht~gas reservoir and \Ht-to-\textsc{Hi} gas ratio over intermediate redshift in \textsc{Hi} flux-limited samples. Our study builds on this analysis and focuses specifically on high stellar mass (M$_\mathrm{*}$ $>$ $10^{10}$ M$_{\odot}$) and high \textsc{Hi} mass (M$_\mathrm{HI}$ $>$ $10^{10.5}$ M$_{\odot}$) samples. 

We analyze the gas fraction over cosmic time, shown in Figure \ref{fig_gf}. The dashed silver line in each panel is the best fit line for the sample of 102 high stellar mass galaxies (M$_{*}$ $>$ $10^{10}$ M$_{\odot}$) that have both \textsc{Hi} and \Ht~measurements (including twenty beyond the local Universe ($z$ $>$ 0.06)). This sample is described in Section \ref{sec_mass} and is shown above the black line in the previous Figure \ref{fig_mass} -- Panel a (solid orange and teal symbols, not including stacked masses). The dashed gold line in the bottom panel is the best fit line for this high stellar mass sample for 29 high \textsc{Hi} mass galaxies with M$_\mathrm{HI}$ $>$ $10^{10.5}$ M$_{\odot}$ (including nine beyond the local Universe). This sample is described in Section \ref{sec_mass} and is shown in the previous Figure \ref{fig_mass} -- Panel b with small gold circles overlaid. The mean cold gas fraction for $z$ $\sim$ 0, 0.25, and 0.5 is shown in Table \ref{tab_gf}, along with the best fit line equations and the standard deviation of the residuals ($\sigma$). The best-fit slopes should be interpreted with caution since there is a large scatter around the fit and a larger sample is required to draw statistically robust results.
		
\begin{table*}			
\caption{Mean Cold Gas Fraction}			
\centering			
\begin{tabular}{lccc}			
\hline			
\hline			
Redshift&	$z$$\sim$0$^{a}$&	$z$$\sim$0.25&	$z$$\sim$0.50\\[1.7ex]
\hline			
M$_\mathrm{HI}$ / M$_{*}$$^{b}$&	0.51 $\pm$ 0.05&	0.34 $\pm$ 0.12&	0.16 $\pm$ 0.23\\
\hspace{0.1cm} y = [-0.70 $\pm$ 0.44] z + [0.51 $\pm$ 0.05], $\sigma$ = 0.34 \hspace{0.1cm} &&&\\			
M$_\mathrm{H2}$ / M$_{*}$$^{b}$&	0.08 $\pm$ 0.02&	0.29 $\pm$ 0.04&	0.50 $\pm$ 0.07\\
\hspace{0.1cm} y = [0.83 $\pm$ 0.14] z + [0.08 $\pm$ 0.02], $\sigma$ = 0.11 \hspace{0.1cm} &&&\\			
M$_\mathrm{HI}$+M$_\mathrm{H2}$ / M$_{*}$$^{b}$&	0.59 $\pm$ 0.05&	0.62 $\pm$ 0.13&	0.65 $\pm$ 0.25\\
\hspace{0.1cm} y = [0.12 $\pm$ 0.50] z + [0.59 $\pm$ 0.05], $\sigma$ = 0.39 \hspace{0.1cm} &&&\\[1.7ex]			
\hline			
M$_\mathrm{H2}$ / M$_\mathrm{HI}$$^{b}$& 	0.20 $\pm$ 0.06&	1.13 $\pm$ 0.15&	2.06 $\pm$ 0.29\\
\hspace{0.1cm} y = [3.72 $\pm$ 0.56] z + [0.20 $\pm$ 0.06], $\sigma$ = 0.44\hspace{0.1cm} &&&\\			
M$_\mathrm{H2}$ / M$_\mathrm{HI}$$^{c}$&     	0.12 $\pm$ 0.11&	0.79 $\pm$ 0.21&	1.46 $\pm$ 0.37\\
\hspace{0.1cm} y = [2.67 $\pm$ 0.70] z + [0.12 $\pm$ 0.11], $\sigma$ = 0.40\hspace{0.1cm} &&&\\[1.7ex]			
\hline			
\hline			
\multicolumn{4}{p{13.5cm}}{\footnotesize{The mean cold gas fractions are derived from the best fit line equations for the galaxy samples. $^{a}$Local Universe ($z$ $<$ 0.06.) $^{b}$M$_{*}$ $>$ $10^{10}$ M$_{\odot}$. $^{c}$M$_\mathrm{HI}$ $>$ $10^{10.5}$ M$_{\odot}$.}}			
\end{tabular}			
\label{tab_gf}			
\end{table*}			

The \textsc{Hi} gas fraction (\textsc{Hi} mass / stellar mass) over redshift is shown in Figure \ref{fig_gf} -- Panel a. There is a decrease in the \textsc{Hi} gas fraction with increasing redshift (1.0$\sigma$), with a mean at our highest redshift that is 3.2 $\pm$ 0.5 times lower than the local Universe mean. Panel b shows the \Ht~gas fraction (\Ht~mass / stellar mass) over redshift. There is an increase in the \Ht~gas fractions with increasing redshift (3.8$\sigma$), with a mean at our highest redshift that is 6.3 $\pm$ 1.8 times higher than the local Universe mean. Panel c shows the total gas fraction (\textsc{Hi} + \Ht~gas mass / stellar mass) over redshift. The total gas fraction is essentially flat throughout the redshift range (0.2$\sigma$), with a mean at our highest redshift that is 1.1 $\pm$ 0.4 times higher than the local Universe mean.

The \Ht~/ \textsc{Hi} gas fraction over redshift is shown in Panel d. \textsc{Hi} makes up the vast majority of the cold gas reservoir in galaxies in the local Universe, with $\sim$ 30\% of the available gas in the \Ht~phase \citep{Catinella2010, Saintonge2011}, shown with a dotted line. There is an increase in the \Ht~/ \textsc{Hi} gas fractions with increasing redshift (4.2$\sigma$), shown with the silver dashed line, for the sample of 102 galaxies with M$_{*}$ $>$ $10^{10}$ M$_{\odot}$. The mean at our highest redshift is 10.3 $\pm$ 3.4 times higher than the local Universe mean. Furthermore, there is an increase with increasing redshift (3.3$\sigma$), shown with the gold dashed line, for the sample of 29 galaxies with the highest \textsc{Hi} mass ($>$ $10^{10.5}$ M$_{\odot}$). The mean at our highest redshift is 12.2 $\pm$ 11.6 times higher than the local Universe mean.

The general consensus from simulations is that the \Ht~/ \textsc{Hi} mass ratio in galaxies evolves with cosmic time, decreasing from high redshift ($z$ $\sim$ 2 -- 3) to the present day ($z$ $\sim$ 0) \citep{Obreschkow2009, Lagos2011, Popping2014, Dave2019, Dave2020}. Although simulations predict the same overall trend, the specific implementation of physical processes (such as the relation to gas surface density, pressure, metallicity, and the interstellar radiation field) can lead to quantitative differences in the exact evolution of the ratio across galaxy populations and environments. Hydrodynamic simulations offer a more physically detailed approach but are limited by computational cost and the need for sub-grid physics models, whereas semi-analytic models use more generalized prescriptions that may not capture all the nuances of galaxy evolution. Furthermore, indirect measurements of \textsc{Hi} via stacking \citep{Bera2019, Chowdhury2020,Chowdhury2021} together with inferred measurements of \Ht~via scaling relations, find an increase in the \Ht~/ \textsc{Hi} ratio from $z$ $\sim$ 1.3 to 1 and a decrease in the ratio from $z$ $\sim$ 1 to 0, for blue star-forming galaxies \citep{Chowdhury2022}. Hence, continued direct measurements of the atomic and molecular content of high redshift galaxies will provide important constraints on these predictions of galaxy evolution across cosmic time.

\subsubsection{Relation to Star Formation Properties}
\begin{figure}
\includegraphics[scale=1.35]{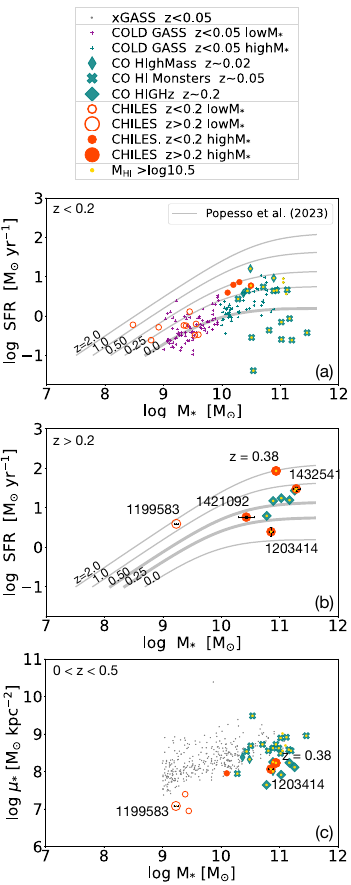}
\caption{
Star formation properties of galaxies. The gray lines show the fit to the star-forming sequences at different redshifts from \citet{Popesso2023}. \textbf{Panel a:} SFR versus stellar mass for galaxies at $z$ $<$ 0.2. The SFMS line for $z$ = 0 is in bold. \textbf{Panel b:} SFR versus stellar mass for galaxies at $z$ $>$ 0.2. The SFMS lines for $z$ = 0.25 and 0.5 are in bold. \textbf{Panel c:} Stellar surface density as a function of stellar mass. The gray circles are $\mu_{*}$ measurements of 352 blue galaxies from the xGASS at $z$ $<$ 0.05.
}
\label{fig_sf}
\end{figure}

The SFMS of galaxies is the correlation between stellar mass and SFR in galaxies and is often used as a measure in studying the evolution of galaxies \citep{Speagle2014, Scoville2017}. This correlation generally shows that larger galaxies have a higher SFR \citep{Renzini2015}. The top two panels in Figure \ref{fig_sf} show the relation of SFR versus stellar mass for galaxies in the \textsc{Hi} sample shown in Figure \ref{fig_mass} at $z$ $<$ 0.2 and $z$ $>$ 0.2. The gray lines show the fit to the star-forming sequences of a comprehensive compilation of observational data from 25 literature studies over a wide range of redshift (0 $<$ $z$ $<$ 6) and stellar mass ($10^{8.5}$ -- $10^{11.5}$ M$_{\odot}$) from \citet{Popesso2023}. The lines from bottom to top are $z$ = 0, 0.25, 0.50, 1, and 2. The assumed dispersion of 0.3 dex on the main-sequence relation is not uniform across all masses and can vary depending on the specific stellar mass range and the time scale over which star formation is averaged.  In Panel a, the majority of the overall sample is from the local Universe ($z$ $<$ 0.06), except for the CHILES sample at $z$ $\sim$ 0.12 (in orange). Most of the high stellar-mass galaxies lie above their SFMS. Two out of three HIghMass galaxies lie around the SFMS $z$ $\sim$ 0.5, along with CHILES high stellar mass galaxies. \textsc{Hi} Monsters lie around SFMS $z$ $\sim$ 0.25 and some far below. xCOLD GASS (high stellar mass) lies around the SFMS $z$ $\sim$ 0 and 0.25. 

In Panel b, a majority of the galaxies lie above the SFMS. Three out of four HIGHz galaxies at $z$ $\sim$ 0.2 lie above SFMS $z$ $\sim$ 0.5. Two CHILES galaxies have significantly higher star formation than is expected at their redshift. The CHILES galaxy at $z$ = 0.38 lies at SFMS $z$ $\sim$ 2. Interestingly, the new galaxy 1199583 at $z$ = 0.29 has 13 times the \textsc{Hi} mass compared to the stellar mass and lies close to SFMS $z$ $\sim$ 2. In contrast, the new galaxy 1203414 at $z$ = 0.26 has an undisturbed stellar component in its JWST image and lies below the $z$ = 0.25 star-forming sequence. The plots show that the large \textsc{Hi} reservoirs of the collective higher-mass samples, particularly beyond the local Universe, are above the SFMS for their stellar mass and redshift. The most \textsc{Hi} rich galaxies also seem to have higher star-formation rates, which may indicate an indirect link between star formation and the \textsc{Hi} gas reservoir. 

Galaxies with large \textsc{Hi} reservoirs are often characterized by their extended disks and lower stellar and gas surface brightness. Panel c shows the stellar mass surface density versus stellar mass.The small gray circles are $\mu_{*}$ measurements of 352 blue galaxies (NUV-r $<$ 3) from xGASS at $z$ $<$ 0.05. In the overall CHILES sample, six galaxies have stellar mass surface density measurements (three from \citet{Hess2025}, one from \citet{Fernandez2016} and two presented in this paper that are shown with stellar mass error bars in Panel b). These galaxies are lower in stellar mass surface density than the local Universe xGASS sample. Galaxy 1203414 (high stellar mass) compares well with the HIGHz sample, and galaxy 1199583 (low stellar mass) compares well with the HIghMass sample. Galaxies 1203414 and 1199583 have extended \textsc{Hi}, 2.3 $\pm$ 0.5 and 3.8 $\pm$ 1.5 times their optical extent, respectively, with overall lower surface densities at fixed stellar mass. Further insight into these parameters could include the characterization of angular momentum and halo spin, along with knowledge of the \textsc{Hi} diameters of other samples.

\subsubsection{Cold Gas Conversion Factor} \label{convf}
Although it is common practice to assume a constant value for the CO conversion factor similar to the Galactic disk, $\alpha_{\text{CO}}$ can vary between galaxies as well as within galaxies, varying by up to orders of magnitude depending on environmental conditions \citep{Bolatto2013}. Departures from the Galactic disk value include galaxy centers \citep{Sandstrom2013, Teng2023}, low-metallicity environments \citep{Leroy2011, Genzel2012, Shi2016}, high star-bursting activity or merging galaxies \citep{Downes1998, MontoyaArroyave2023}, outflows \citep{Pereira-Santaella2024}, and galaxies at high redshift \citep{daCunha2013}. There has been significant effort to establish a robust prescription for the conversion function. \citet{Chiang2024} examine how $\alpha_{\text{CO}}$ scales with several physical quantities, finding that integrated CO intensity, interstellar radiation field strength, and star formation rate surface density have the strongest correlations with $\alpha_{\text{CO}}$, using resolved measurements of 37 nearby galaxies. In addition, with $\sim$610 independent measurements of CO(1--0), they report a mean value of 4.2 for $\alpha_{\text{CO}}$. In a recent study using a cross-calibrated sample of 407 metal-rich and/or massive galaxies spanning $z$ = 0 -- 5 and more than four orders of magnitude in infrared luminosity, \citet{Dunne2022} suggest that a value of $\alpha_{\text{CO}}$ = 4.0 (0.14 dex) is adequate for global molecular gas estimates. The xCOLD GASS sample uses a multivariate conversion function \citep{Accurso2017} that depends primarily on metallicity and secondarily on the SFMS offset, a parameter related to the field strength of UV radiation. Their $\alpha_{\text{CO}}$ vary between 1.0 -- 24.5, with an average of 4.4 (Table \ref{tab_samples}). Of the 29 galaxies in the high \textsc{Hi} mass sample ($>$ $10^{10.5}$ M$_{\odot}$), 20 are above the SFMS for their stellar mass and seven have SFR that are an order of magnitude higher than the Milky Way. Lower conversion factors relating to star-formation applied across this sample may change the y-intercept of the line more than the slope of the line itself. More studies are needed that contain combinations of the full parameter space of factors that can depart from the Galactic disk $\alpha_{\text{CO}}$ for large samples of galaxies.

\begin{figure}
\includegraphics[scale=1.45]{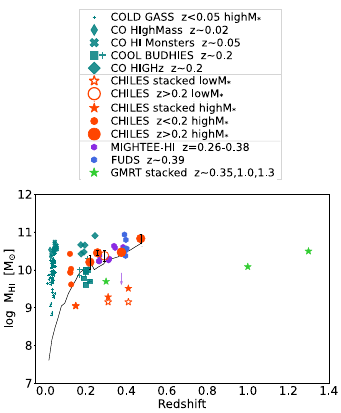}
\caption{
\textsc{Hi} in galaxies over extended redshift. The plot includes the high stellar mass sample above the line in Figure \ref{fig_mass} -- Panel a (minus the galaxies with the thick teal crosses and plus the newly presented galaxy with the large open orange circle), the low and high mass stacked CHILES results, plus three added samples. The purple hexagons and light purple arrow are from the MIGHTEE-HI survey at $z$ = 0.26 -- 0.38. The blue hexagons are from the FUDS survey at $z$ $\sim$ 0.39. The green stars are from the stacked GMRT results at $z$ $\sim$ 0.35, 1.0, and 1.3. The solid black line is the CHILES \textsc{Hi} mass detection limit shown in Figure \ref{fig_mass} -- Panel b. 
}
\label{fig_plus}
\end{figure}

\subsubsection{\textsc{Hi} $\mathrm{beyond}$ $z$ $\mathrm{= 0.5}$} \label{beyond}
The first reliable statistical results on the average \textsc{Hi} content of selected samples of galaxies beyond $z$ = 0.5 have been obtained with the GMRT in a series of stacking experiments.  While \citet{Bera2019} find that there is no evolution up to $z$ = 0.4, \citet{Chowdhury2020,Chowdhury2021} find that there is true evolution in the \textsc{Hi} content for star-forming galaxies between $z$ = 1 and $z$ = 1.3. Several surveys are now also publishing the first direct detection in \textsc{Hi} of individual galaxies at redshifts larger than 0.2. In Figure 14, we add \textsc{Hi} samples from MIGHTEE-HI \citep{Jarvis2025} with \textsc{Hi} masses ranging from 1.6 -- $4.3\times10^{10}$ M$_{\odot}$ and FUDS \citep{Xi2024} with \textsc{Hi} masses ranging from 2.3 -- $8.5\times10^{10}$ M$_{\odot}$. We also add for comparison the stacking results of the GMRT and the CHILES result of \citet{Luber2025a}. The line indicates the sensitivity limit of CHILES. All direct detections, including MIGHTEE-HI, are just above this sensitivity limit. So far, the direct detections are probing the tip of the iceberg and cannot be used to make general statements about the \textsc{Hi} gas content evolution beyond $z$ = 0.2. They are, however, fascinating systems to study individually.

\section{Summary}
1) Images -- The CHILES survey uses the VLA to image \textsc{Hi} over a continuous redshift range 0 $<$ $z$ $<$ 0.5 with 856 hours of observation at a single pointing in the COSMOS field. In this high-redshift analysis, we present for the first time resolved distribution, morphology, plus kinematics of the \textsc{Hi} gas in four galaxies in the redshift range of $z$ = 0.22 to 0.47, with a resolution of 22 to 43 kpc (6.3$^{\prime\prime}$ to 7.5$^{\prime\prime}$), respectively. 

\begin{itemize}
    \item Galaxy 1432541 ($z$ = 0.47) is a massive system that has \textsc{Hi} extending nearly twice its asymmetric optical extent. JWST imaging reveals a compact knot and this galaxy lies far above its SFMS (at $z$ $\sim$ 0.8). It has both \Ht~molecular and [OII] ionized gas measurements redshifted from the optical and \textsc{Hi}, revealing the complex gas kinematics of this system. 

    \item Galaxy 1199583 ($z$ = 0.29) is an irregular low-mass dwarf-like system that has a huge \textsc{Hi} envelope (13$\times$ stellar mass) extending nearly four times its asymmetric optical extent. JWST imaging also shows a compact knot, and this galaxy lies far above the SFMS (at $z$ $\sim$ 1.7) with no continuum and no \Ht~detection.

    \item Galaxy 1203414 ($z$ = 0.26) is a large optically symmetric spiral that was easily found visually and has asymmetric \textsc{Hi} extending more than two times its optical extent. The galaxy lies below its SFMS (at $z$ $\sim$ 0.15) and the optical, \textsc{Hi} and \Ht~line centers align in this galaxy.

    \item Galaxy 1421092 ($z$ = 0.22) is a large optically symmetric spiral that has asymmetric \textsc{Hi} extending close to one and a half times its optical extent. Similarly to galaxy 1432541, this galaxy has \Ht~gas that is offset (redshifted) from the optical and HI.
\end{itemize}

The four galaxies consist of three high stellar mass systems (with nearby neighbors) and one low stellar mass (isolated) system. They have extended \textsc{Hi}, are actively forming stars and have more \textsc{Hi} than \Ht~gas.\\

2) Properties -- We present \Ht~measurements from the LMT of the four galaxies for a combined look at the cold gas properties. We combine these galaxies into a total CHILES sample of 19 galaxies beyond the local Universe and compare them to galaxy samples at other redshifts. This provides, for the first time, a continuous look at directly detected \textsc{Hi} and \Ht~in emission in individual galaxies over the redshift range 0 $<$ $z$ $<$ 0.5. All galaxy samples have been adjusted to use a common CO-to-\Ht~conversion factor. We investigate where our \textsc{Hi}-selected sample of galaxies falls onto the scaling relations.
\begin{itemize}
    \item \textsc{Hi} selected galaxies beyond the local universe (z $>$ 0.2) appear to have lower stellar surface brightness and are typically above the SFMS for their stellar mass and redshift.

    \item Of the sample of galaxies with \textsc{Hi} mass larger than $>$ $10^{10.5}$ M$_{\odot}$, 20 of 29 are forming stars at a rate that places them well above the SFMS.

    \item For a high stellar mass sample (M$_{*}$ $>$ $10^{10}$ M$_{\odot}$) of 102 galaxies: the mean \textsc{Hi} gas fraction at our highest redshift is 3.2 $\pm$ 0.5 times lower than the local Universe mean; the mean \Ht~gas fraction at our highest redshift is 6.3 $\pm$ 1.8 times higher than the local Universe mean; the mean total gas fraction at our highest redshift is 1.1 $\pm$ 0.4 times higher than the local Universe mean; the mean \Ht~/ \textsc{Hi} gas fraction at our highest redshift is 10.3 $\pm$ 3.4 times higher than the local Universe mean.

    \item For a high \textsc{Hi} mass sample (M$_\mathrm{HI}$ $>$ $10^{10.5}$ M$_{\odot}$) of 29 galaxies, the mean \Ht~/ \textsc{Hi} gas fraction at our highest redshift is 12.2 $\pm$ 11.6 times higher than the local Universe mean.
\end{itemize}

This analysis with only a few galaxies in an HI-biased sample is a step forward in assessing the evolution of cold gas content with redshift, but a larger sample size with a complete parameter space is required to fully clarify whether these trends hold or are a consequence of low-number statistics.

\section{Acknowledgments} \label{sec:acknowledgments}

We thank the referee for very thorough and thoughtful comments that helped enhance the paper. We thank W. M. Goss for helpful comments on the paper. We thank Moses Mogotsi for guidance and expertise on the SALT DDT. We thank Patrick Kamieneski (plus Jake Summers) for guidance in processing JWST NIRCam images. We thank K. Vinsen and E. da Cunha who ran the SED fitting on the G10/COSMOS v05 catalog with MAGPHYS. JBB was a Jansky Fellow of the National Radio Astronomy Observatory. The National Radio Astronomy Observatory is a facility of the National Science Foundation operated under cooperative agreement by Associated Universities, Inc. The CHILES survey was partially supported by a collaborative research grant from the National Science Foundation under grant Nos. AST—1412843, 1412578, 1413102, 1413099, and 1412503. Support for this work is also provided by the NSF through award SOSP 18$\_$3133 from the National Radio Astronomy Observatory (NRAO). DJP greatly acknowledges support from the South African Research Chairs Initiative of the Department of Science and Technology and National Research Foundation. The data described in this paper include LMT observations conducted under the scientific program 2023-S1-US-18. The LMT welcomes acknowledgment of the scientific and technical support offered by the LMT staff during the observations and generation of data products provided to the authors. This work would not have been possible without long-term financial support from the Mexican Humanities, Science and Technology Funding Agency, CONAHCYT (Consejo Nacional de Humanidades, Ciencias y Tecnologias), and the US National Science Foundation (NSF), as well as the Instituto Nacional de Astrofisica, Optica y Electronica (INAOE) and the University of Massachusetts, Amherst (UMass). The operation of the LMT is currently funded by CONAHCYT grant \#297324 and NSF grant \#2034318. Some of the observations reported in this paper were obtained with the SALT teleescope [2022-2-DDT-003, PI: J. Blue Bird, CoPIs: M. Mogotsi, D.J. Pisano].

Facilities: VLA, LMT, HST, JWST;
Software: astropy, CASA, CARTA

\bibliography{refs}


\begin{sidewaystable}[]									
\centering									
\caption{Sample Properties}									
\begin{tabular}{l||c|ccc|ccc|cc}									
\hline									
\hline									
\textbf{\textsc{Hi}:}&	Maddox&	xGASS&	HIghMass&	\textsc{Hi} Monsters&	HIGHz&	BUDHIES&	CHILES&	CHILES stack&	GMRT stack\\
&	[Brown Line]&	[Sm Gray Circle]&	&	&	&	&	[Pink Circle]&	[Orange Star]&	[Green Star]\\[1.7ex]															
\hline									
Observation&	ALFALFA&	Arecibo&	ALFALFA&	ALFALFA&	Arecibo&	WSRT&	VLA&	VLA&	GMRT\\
Selection [M$_{\odot}$]&	M$_\textsc{Hi}$$>$$10^{7.5}$&	M$_{*}$$>$$10^{9-11.5}$&	M$_\textsc{Hi}$$>$$10^{10}$&	M$_\textsc{Hi}$$>$$10^{10.5}$&	M$_{*}$,M$_\textsc{Hi}$$>$$10^{10}$&	Clusters&	Untargeted &	--&	--\\																				Count&	9153&	1179&	34&	20&	39&	166&	31&	6&	3\\
\hline									
Redshift&	 $<$0.06&	 $<$0.05&	 $<$0.06&	 0.04--0.08&	 0.17--0.25&	  0.16--0.22&	 0--0.5&	0.15,0.31,0.41&	 0.3,1.0,1.3\\
Environment&	Various&	Various&	Isolated&	Various&	Isolated&	Clusters&	Various &	&	\\
\hline									
M$_{*}$ Calculated&	*ugriz&	*ugriz&	*ugriz&	*ugriz&	*ugriz&	FUV--FIR&	FUV--FIR&	FUV--FIR&	K--band(NIR)\\
SFR Calculated&	--&	NUV,MIR&	*ugriz,H$\alpha$&	*ugriz&	*ugriz&	--&	FUV--FIR&	FUV--FIR&	1.4 GHz Cont\\
Color Cut&	--&	NUV--r$<$3&	--&	--&	NUV--r$<$3&	--&	NUV--r$<$3&	NUV--r$<$3&	NUV--r$<$3\\[1.7ex]
\hline									
\hline									
\textbf{\textsc{Hi} + \Ht:}&	&	xCOLD GASS&	CO HIghMass&	CO \textsc{Hi} Monsters&	CO HIGHz&	 COOL BUDHIES&	CO CHILES&	&	\\
&	&	[Small Cross]&	[Teal Slim Diamond]&	[Teal X]&	[Teal Diamond]&	[Teal Box]&	[Orange Circle]&	&	\\[1.7ex]				\hline									
Observation&	&	IRAM&	IRAM&	FCRAO&	ALMA&	LMT&	LMT,ALMA&	&	\\
Count&	&	532&	3&	20&	5&	23&	19&	&	\\
Count Included&	&	82 L, 65 H&	3&	15&	5&	6 det, 6 UL&	9 det, 10 UL&	&	\\
\hline									
CO Line&	&	CO(1--0)&	CO(1--0)&	CO(1--0)&	CO(1--0)&	CO(1--0)&	CO(1--0)&	&	\\
$\alpha_{\text{CO}}$&	&	4.4 average&	 **3.2&	 **3.2&	**3.2&	4.6&	4.3&	&	\\
   Helium (He)&	&	He included&	He not incl.&	He not incl.&	He not incl.&	He included&	He included&	&	\\[1.7ex]
\hline									
\hline									
\multicolumn{10}{p{16cm}}{\footnotesize{*MPA--JHU catalog based on SDSS DR7. **Values using the CO to  \Ht~conversion factor $\alpha_{\text{CO}}$ of 3.2 are adjusted to 4.3 in the plots.}}									
\end{tabular}									
\label{tab_samples}									
\end{sidewaystable}																		
\end{CJK*}
\end{document}